\documentclass{article}
\usepackage{amsmath}
\usepackage{amssymb}
\usepackage{amsfonts}
\usepackage{graphicx}
\def\beq{\begin{equation}}
\def\eeq{\end{equation}}

\begin{document}
\title{Wavelets in Field Theory}
\author{Fatih Bulut\\
In\"on\"u University\\
Wayne Polyzou\\
The University of Iowa\footnote{This work supported                             
in part by the U.S. Department of Energy, under contract DE-FG02-86ER40286}}

\maketitle

\begin{abstract}

  We advocate the use of Daubechies wavelets as a basis
  for treating a variety of problems in quantum field theory.  This
  basis has both natural large-volume and short-distance cutoffs, has
  natural partitions of unity, and the basis functions are all related
  to the fixed point of a linear renormalization group equation.

\end{abstract}


\section{Introduction}

Daubechies wavelets and their associated scaling functions are
functions \cite{daubechies:1988} \cite{daubechies}
\cite{kaiser}\cite{resnikoff}\cite{jorgensen}\cite{wavelets} that are
used in signal processing and data compression.  The functions have
useful properties that have advantages for applications to problems in
quantum field theory.  Daubechies wavelets have been used in some
field-theoretic applications \cite{best:1994} \cite{best:2000}
\cite{Ismail1:2003} \cite{Ismail2:2003} and different types of
wavelets have been used in more formal treatments of field theory
\cite{federbush:1995} \cite{Battle:1999} \cite{altaisky:2007}
\cite{albeverio:2009} \cite{altaisky:2010} however many useful
properties of wavelets do not appear to have been exploited.  Some of
the properties of Daubechies wavelets that are potentially useful for
field theory applications are:

\begin{itemize}

\item[1.] The functions are an orthonormal basis for the square integrable
functions on the line.  This means that one can formally express fields
{\it exactly} as infinite linear combinations of discrete field
operators and coefficient functions.  This can be used to study the nature 
of corrections to truncated theories or to compute the singular
coefficients in operator product expansions.

\item[2.] The basis functions have compact support. This means 
that discrete fields associated with basis functions with space-like
separated support commute.  The maximum size of the support of the
basis functions can be controlled.  There are also infinite numbers of
basis functions with arbitrarily small support so it is possible to use
these operators study locality and how locality is violated in
truncations.  In these representations local fields can be replaced
by a local algebra of well-defined operators.

\item[3.] The Daubechies basis functions are related to solutions of a
linear renormalization group equation.  It is a natural basis to
formulate renormalization group transformations.  The basis allows for
a separation of scales; a fine resolution truncated Lagrangian or
Hamiltonian can be exactly expressed in terms of a coarse resolution
truncated Lagrangian or Hamiltonian of the same form, plus some
addition operators that fill in the missing fine-scale physics.
Eliminating the fine-scale operators gives a new Hamiltonian,
involving the same masses and coupling constants, in the coarse-scale
degrees of freedom that includes the effects of the eliminated 
fine-scale degree of freedom.  This can be repeated to include the effects
of eliminated degrees of freedom on arbitrarily small scales.

\item[4.] The basis is generated from a single function, which is the
solution of the linear renormalization group equation, using discrete
unitary translations and discrete unitary scale transformations.
While the basis functions have a fractal structure, overlap integrals
of products of any number of the basis functions and low-order
derivatives can be computed analytically using only the
renormalization group equation and a scale fixing condition.  In
addition, integrals of products of basis functions and polynomials can
also be computed exactly using the renormalization group equations.

\item[5.] Subsets of the basis functions contain compact, locally-finite 
partitions of unity.  This property can be used to investigate
symmetries like Poincar\'e invariance.  The partitions of unity can be
used to make local realizations of the Lie algebra.  The generators
can be expressed exactly as sums of localized operators obtained by
integrating the operator densities multiplied by the functions in the
partition of unity.

\item[6.] The basis functions have a limited amount of smoothness, and
can locally pointwise represent low-degree polynomials.  This means
that overlaps of low-order derivatives of field operators can be
computed directly, rather than by using a finite-difference 
approximation.
 
\item[7.] Because the basis functions have compact support and there are
basis functions with arbitrarily small support, the basis has both 
natural short-distance and large-volume cutoffs.

\end{itemize} 

The purpose of this paper is to discuss some problems in field theory
where wavelets might have advantages as well as some of the
specialized methods for dealing with fractal basis functions.  In
section two we discuss the construction of the wavelet basis and
introduce the renormalization group equation for the Daubechies
scaling function.  In section three we introduce our free-field
conventions.  In section four we introduce the wavelet representation
of free fields.  This representation is also relevant in
interacting theories, when the natural volume and resolution
truncations in the wavelet representation are used to make finite
number of degree-of-freedom truncations of the interacting theory.
In section five we discuss the scaling properties of the wavelet
fields.  In section six we discuss Poincar\'e invariance and
partitions of unity.  In section seven we discuss the formulation of
renormalization group equations in the wavelet basis.  In section
eight we discuss gauge transformations in the wavelet representation.
In the appendix we compute the overlap coefficients that couple
different scales in the free-field Hamiltonian.
  
\section{Basis construction}

The Daubechies scaling function is the solution of a linear
renormalization group equation subject to a scale-fixing normalization
condition.  All basis functions are generated from the scaling
function using discrete translations, dyadic scale transformations,
and sums.

The discrete translation operator, $T$, and the dyadic scale
transformation, $D$, are unitary operators on $L^2(\mathbb{R})$ defined
by
\beq
(D f)(x) = \sqrt{2} f(2x) 
\qquad
(Tf)(x) = f(x-1) .
\label{A:1}
\eeq
The operator $D$ decreases the support of a function with compact
support by a factor of two in a manner that preserves the
$L^2(\mathbb{R})$ norm.  The scaling function, $s (x)$, is the
solution of the linear renormalization group equation,
\beq
s (x) = \sum_{n=0}^{2K-1} h_n D T^n s (x) ,
\label{A:2}
\eeq
called the scaling equation.
Equation (\ref{A:2}) states that the scaling function is a linear
combination of $2K$-translated copies of itself on a scale smaller by
a factor of 2.  Since this is a homogeneous equation for $s(x)$, we
are free to assign the normalization.  It is fixed by the scale-fixing
condition
\beq
\int s (x) dx =1.
\label{A:3}
\eeq
Equations (\ref{A:2}-\ref{A:3}) define the scaling function.  
The $h_n$ are real
coefficients that are characteristic of the type of wavelet. 
The coefficients for some of the Daubechies scaling functions are given 
in Table 1.

Given the solution of (\ref{A:2}-\ref{A:3}), scale-$1/2^k$ 
scaling functions are defined 
by applying $n$ unit translations followed by $k$ dyadic scale
transformations to the original scaling function,
\beq
s_n^k (x) := D^k T^n s (x) .
\label{A:4}
\eeq  
Linear combinations
of these functions with square-sumable coefficients span a subspace
${\cal H}_k$ of $L^2(\mathbb{R})$, which we call the
resolution-$1/2^k$ subspace.  The scaling equation implies the relation
\beq
{\cal H}_{k+1}\supset {\cal H}_{k}
\eeq
and more generally 
\beq
{\cal H}_{k+m}\supset {\cal H}_{k}
\eeq 
for $m>0$.  This means that the lower-resolution spaces are linear 
subspaces of the higher-resolution spaces.

The mother wavelet, $w (x)$, is defined as the following linear
combination of the $s_n^1(x)$ scaling functions:
\beq
w (x)  = \sum_{n=0}^{2K-1} g_n D T^n s(x) =
\sum_{n=0}^{2K-1} g_n s^1_n (x) 
\label{A:5}
\eeq
where the coefficients $g_n$ are related to the coefficients 
$h_n$ by reversing the order and alternating the signs
\beq
g_n = (-)^n h_{2K-1-n} .
\label{A:6}
\eeq
Plots of the Daubechies $K=3$ scaling function and mother wavelet 
are given in figures 1 and 2.
\begin{figure}
\begin{center}
\includegraphics[width=10.0cm]{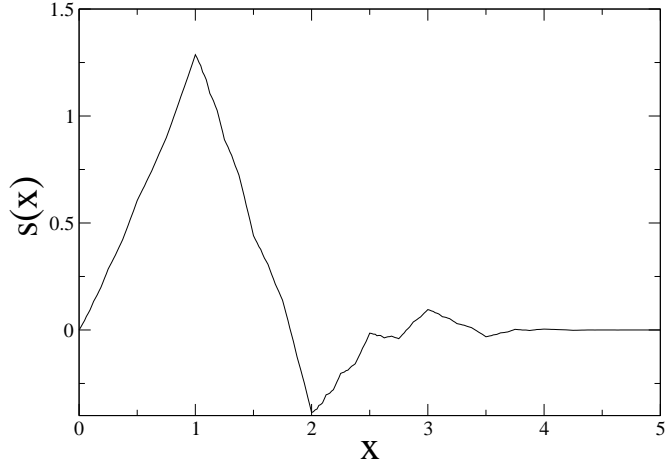}
\caption[Short caption for figure 1]{\label{labelFig1} 
Daubechies $K=3$ scaling function.}
\end{center}
\label{fig.1}
\end{figure}
\begin{figure}
\begin{center}
\includegraphics[width=10.0cm]{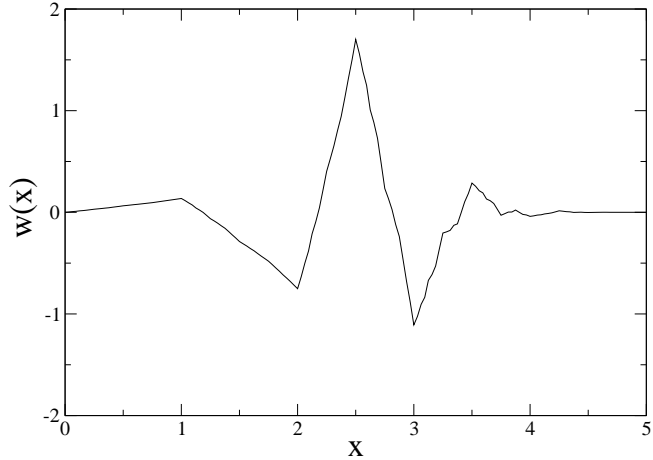}
\caption[Short caption for figure 2]{\label{labelFig2} Daubechies $K=3$ 
mother wavelet}
\end{center}
\label{fig.2}
\end{figure}
Scale-$1/2^k$ wavelets, $w_n^k(x)$, are constructed 
by applying translations and dyadic scale transformations to the 
mother wavelet
\beq
w_n^k (x) = D^k T^n w (x) .
\label{A:7}
\eeq
The scale $1/2^k$ wavelets are designed to span the orthogonal complement, 
${\cal W}_k$, of ${\cal H}_{k}$ in ${\cal H}_{k+1}$,
\beq
{\cal H}_{k+1} = {\cal H}_k \oplus {\cal W}_k .
\label{A:8}
\eeq
The coefficients $h_l$ are chosen so the translates of the scaling
function and mother wavelet are orthonormal and finite linear
combinations of translations of the scaling function can be used to
make local pointwise representations of polynomials of degree $K-1$.

Continuing the decomposition (\ref{A:8}) inductively leads to the following
orthogonal direct sum decomposition of the Hilbert space of square
integrable functions on the line
\beq
L^2(\mathbb{R})= {\cal H}_k \oplus {\cal W}_k 
\oplus {\cal W}_{k+1} \oplus \cdots .
\label{A:8a}
\eeq  
This gives a decomposition of the Hilbert space into orthogonal subspaces 
with successively finer resolutions.  Specifically, the 
basis for ${\cal H}_k$ can be used to represent features down to scale 
$1/2^k$, ${\cal W}_k$ represents features down to scale $1/2^{k+1}$ that 
cannot be represented on scale $1/2^k$, 
${\cal W}_{k+1}$ represents features down to scale $1/2^{k+2}$ that 
cannot be represented on scale $1/2^{k+1}$, etc..  
Smearing local fields with these basis functions gives discrete fields 
that are sensitive to average information on different scales. 

The coefficients $h_n$, which characterize the Daubechies $K$-wavelets, 
are solutions of the system of equations
\beq
\sum_{n=0}^{2K-1} h_n = \sqrt{2}
\label{A:9}
\eeq
\beq
\sum_{n=0}^{2K-1} h_n h_{n-2m} = \delta_{m0}
\label{A:10}
\eeq
\beq
\sum_{n=0}^{2K-1} n^m g_n =
\sum_{n=0}^{2K-1} n^m (-1)^n h_{2K-1-n} = 0
\qquad m < K . 
\label{A:11}
\eeq
The first equation is a necessary condition for the scaling equation
to have a solution; it can be derived by taking the Fourier transform of
both sides of the scaling equation.  The second equation ensures that
integer translates of the scaling function are orthonormal.  The third
equation ensures that linear combinations of integer translates of the
wavelet functions are orthogonal to degree $K-1$ polynomials.

This last condition, along with (\ref{A:8a}), implies that 
polynomials of degree $K-1$ can be locally pointwise represented by 
locally finite linear combinations of scaling functions. 

The solution of these equations for $K=1,2,3$ are given in Table 1.  
The solutions are unique up to reversal of order; $n \to  2K-1-n$.
\begin{table} 
\centering
\begin{tabular}{|l|l|l|l|}
\hline
$h_n$ & K=1 & K=2 & K=3   \\
\hline					      		      
$h_0$ &$1 / \sqrt{2}$  & $(1+\sqrt{3})/4\sqrt{2}$ &$(1+\sqrt{10}+\sqrt{5+2\sqrt{10}})/16\sqrt{2}$ \\
$h_1$ &$1 / \sqrt{2}$  & $(3+\sqrt{3})/4\sqrt{2}$ & $(5+\sqrt{10}+3\sqrt{5+2\sqrt{10}})/16\sqrt{2}$ \\
$h_2$ &$0$  & $(3-\sqrt{3})/4\sqrt{2}$ & $(10-2\sqrt{10}+2\sqrt{5+2\sqrt{10}})/16\sqrt{2}$ \\
$h_3$ &$0$  & $(1-\sqrt{3})/4\sqrt{2}$ & $ (10-2\sqrt{10}-2\sqrt{5+2\sqrt{10}})/16\sqrt{2} $ \\
$h_4$ &$0$  & $0$ & $(5+\sqrt{10}-3\sqrt{5+2\sqrt{10}})/16\sqrt{2}$ \\
$h_5$ &$0$  & $0$ & $(1+\sqrt{10}-\sqrt{5+2\sqrt{10}})/16\sqrt{2}$ \\
\hline
\end{tabular}
\caption{}
\end{table}
The numerical values of the $K=3$ coefficients, rounded to two decimal
places, are $h_0 = - g_5 =.43$, $h_1 = g_4 =1.11$, $h_2 = - g_3 =.66$,
$h_3 = g_2 = -.34$, $h_4 = - g_1 =-.39$, $h_5 = g_0 =-.66$.  An
important observation is that only one of the six coefficients,
$h_1=g_4$  has magnitude larger than $1$.

In this work we advocate using the Daubechies $K=3$ wavelets.  This is
because they have small support and have one continuous derivative.
The derivative allows one to replace the finite-difference
approximations of derivatives that appear in the Hamiltonian or
Lagrangian densities with actual derivatives.

The scaling function $s(x)$ and mother wavelet $w(x)$
have support on the interval $[0,2K-1]$.  It follows from 
(\ref{A:8a}) that for any fixed scale, $1/2^k$, the functions 
\beq
\{ s^k_n(x)\}_{n=-\infty}^{\infty} \cup 
\{ w^m_n(x)\}_{n=-\infty, m=k}^{\infty,\infty}  
\label{A:12}
\eeq
are an orthonormal basis for $L^2(\mathbb{R})$.  The basis functions
$s^k_n (x)$ and $w^k_n (x)$ have compact support on the interval
$[2^{-k}n, 2^{-k}(n+2K-1)]$.  This support justifies calling
${\cal H}_k$ the resolution-$1/2^k$ subspace.

The basis functions satisfy the orthonormality conditions
\beq
\int dx s^k_n (x) s^k_m (x) = \delta_{mn}
\label{A:12a}
\eeq
\beq
\int dx s^k_n (x) w^{k+l}_m (x) = 0 \qquad (l\geq 0)
\label{A:12b}
\eeq
\beq
\int dx  w^{k}_n (x) w^{l}_m (x) = \delta_{mn} \delta_{kl}.
\label{A:12c}
\eeq
Scaling functions with {\it different} $k$ are not
orthogonal and are not members of the same basis.  Similarly, the
$w^{k+l}_m (x)$ with negative $l$ are not orthogonal to $s^k_m(x)$ and
are not members of the basis that includes the $s^k_m(x)$.  In
general, the basis (\ref{A:12}) consists of scaling functions,
$s_m^k(x)$, of a
fixed largest scale, $1/2^k$, and wavelets, $w^{k+l}_m(x)$, 
that can represent the
structure on all scales, $1/2^{k+l}$, with $l\geq 0$.

A basis for the resolution $1/2^{k+m}$  subspace, ${\cal H}_{k+m}$, that
includes the basis functions for the resolution $1/2^k$ subspace, ${\cal
H}_{k}$, is
\beq
\mbox{span}(\{ s^{k+m}_n\}_{n=-\infty}^{\infty}) = 
\mbox{span}( \{ s^k_n\}_{n=-\infty}^{\infty} \cup 
\{ w^l_n\}_{n=-\infty, l=k}^{\infty,k+m-1}).
\label{A:13}
\eeq
This means that the wavelets on scales $1/2^k,1/2^{k+1},\cdots, 1/2^{k+m-1}$ 
fill in all of the missing finer-resolution information in 
${\cal H}_{k+m}$ that is not contained in ${\cal H}_{k}$. 

The scaling functions form a partition of unity
\beq
1 = \sum_{n=-\infty}^{\infty} s (x-n) = 
\sum_{n=-\infty}^{\infty} (T^n s) (x) = \sum_{n=-\infty}^{\infty} s_n^0 (x).  
\label{A:14}
\eeq
This partition of unity is locally finite, meaning that at any point
only a finite number of the functions are non-zero, and the 
functions all have compact support.
There are also partitions of unity for the scale-$1/2^k$ scaling functions
\beq
\sum_{n=-\infty}^{\infty} { 1 \over \sqrt{2^k}} s^k_n (x) = 
\sum_{n=-\infty}^{\infty} s (2^k x -n) = 1  
\label{A:15}
\eeq
except they must be multiplied by the factor 
${ 1 \over \sqrt{2^k}}$.

The scale-$1/2^{k-1}$ scaling functions and scale-$1/2^{k-1}$ wavelets
are both linear combinations of the scale $1/2^k$ scaling functions
\beq
s^{k-1}_n (x)   = \sum_{l=0}^{2K-1} h_l  s^k_{2n+l} (x)
\label{A:16}
\eeq
\beq
w^{k-1}_n (x)   = \sum_{l=0}^{2K-1} g_l s^k_{2n+l} (x) .
\label{A:17}
\eeq
The inverse of these relations express the scale $1/2^k$ scaling function
as a linear combination of the scale $1/2^{k-1}$ scaling functions and 
wavelets
\beq
s^k_n (x)= \sum_m h_{n-2m} s^{k-1}_m(x) +  \sum_m g_{n-2m} w^{k-1}_m(x) .
\label{A:18}
\eeq
Equations (\ref{A:16}-\ref{A:17}) define the wavelet transform in
signal processing.  It is a pair of filters that decompose a signal
into high and low-resolution parts in a manner that can be inverted to
reconstruct the full signal.  In signal processing applications it is
applied many times to generate a multi-resolution decomposition of the
signal.  Equation (\ref{A:18}) is relevant in field-theory
applications, where we see that the wavelets represent the lost
information when one discards small-scale degrees of freedom.  The same
coefficients appear in (\ref{A:16}-\ref{A:17}) and (\ref{A:18})
because these two sets of orthonormal basis functions on ${\cal H}^k$
are related by a real orthogonal transformation.

For some applications it is useful to express equations (\ref{A:16}) and
(\ref{A:17}) in infinite matrix form
\beq
s^{k-1}_n (x)   = \sum_{m=2n}^{2n+2K-1} H_{nm}  s^k_{m} (x)
\label{A:19}
\eeq
\beq
w^{k-1}_n (x)   = \sum_{m=2n}^{2n+2K-1} G_{nm} s^k_{m} (x)
\label{A:20}
\eeq
where
\beq
H_{nm}= h_{m-2n} \qquad
G_{nm}= g_{m-2n} . 
\label{A:22}
\eeq
In this notation (\ref{A:18}) becomes 
\beq
s^k_n(x) = \sum_m {H}^t_{nm} s^{k-1}_m(x) +  \sum_m {G}^t_{nm} 
w^{k-1}_m(x)
\label{A:21}
\eeq
where 
\beq
{H}^t_{nm}=H_{mn}=  h_{n-2m} \qquad 
{G}^t_{nm}=G_{mn}=  g_{n-2m} .
\label{A:23}
\eeq
While the matrices are in principle infinite, for fixed $n$ 
there are only $2K$ non-vanishing terms in the sum, which
is clear in the equivalent expressions (\ref{A:16}) and
(\ref{A:17}).

Note that if we set $x=n$ in the scaling equation it becomes 
\beq
s (n) = \sum_{l=0}^{2K-1} \sqrt{2} h_l  s (2n-l)
= \sqrt{2} \sum_{m=2n-2K+1}^{2n} h_{2n-m}  s (m)
\label{A:24}
\eeq
which has the form of an eigenvalue equation for the matrix,
$M_{nm}= h_{2n-m}$, where the eigenvectors are 
$(0, s(1), s(2) , \cdots , s(2K-2),0)$ and the eigenvalue is
$1/\sqrt{2}$.  If we differentiate the scaling equation and again
evaluate the derivative of the scaling functions at integer values 
we get    
\beq
s' (n) = \sum_{l=0}^{2K-1} 2\sqrt{2} h_l  s' (2n-l)
= 2\sqrt{2} \sum_{m=2n-2K+1}^{2n} h_{2n-m}  s' (m) .
\label{A:25}
\eeq
This is an eigenvalue equation for the {\it same matrix}, with eigenvector
$(s'(0), s'(1)$, $s'(2) , \cdots , s'(2K-2),s'(2K-1))$ and eigenvalue
$1/2^{3/2}$.  Each derivative brings down an additional
factor of 1/2 in the eigenvalue.  Since a finite matrix has a
finite number of eigenvalues, the scaling function can only be
differentiated a finite number of times, ($n$ times if the matrix
above has an eigenvalues of $2^{-(n+1/2)}$).  Thus we have the
interesting property that the number of derivatives is associated with
the eigenvalue spectrum of a finite-dimensional matrix.  This makes it
clear that the individual wavelets and scaling functions have only a
finite number of derivatives.  It is interesting that certain linear
combinations of scaling functions may have more derivatives than the
functions themselves; this is the case with the partition of unity
(\ref{A:14}).  The sum is constant, which is infinitely
differentiable, but the individual functions in the sum are not!

For fields that are functions of four space-time variables, we use a
basis for the square integrable functions of the three space variables
consisting of products of the wavelet basis functions, (\ref{A:12}),
in each of the three variables.  We identify basis functions that are
the product of three resolution $1/2^k$ scaling functions as the
resolution $1/2^k$ basis; the seven other products of combinations of
wavelets and scaling functions span the orthogonal complement of the
resolution $1/2^k$ subspace in $L^2(\mathbb{R}^3)$.  We introduce a
compact notation to make the three-dimensional case look like the
one-dimensional case.

In three dimensions we define: 
\beq
s_{\mathbf{n}}^{k} (\mathbf{x}) = 
s^k_{n_1} (x_1) s^k_{n_2} (x_2) s^k_{n_3} (x_3) 
\label{A:26}
\eeq
and use the notation to $w_{\mathbf{n},\alpha}^{m} (\mathbf{x})$ to
represent the remaining basis functions that have one of the following
seven forms
\beq
w_{\mathbf{n},1,k_3 }^{m} (\mathbf{x}) := s^k_{n_1} (x_1) s^k_{n_2} (x_2) w^{k_3}_{n_3} (x_3) 
\label{A:27}
\eeq
\beq
w_{\mathbf{n},2,k_2}^{m} (\mathbf{x}) := s^k_{n_1} (x_1) w^{k_2}_{n_2} (x_2) s^k_{n_3} (x_3) 
\label{A:28}
\eeq
\beq
w_{\mathbf{n},3,k_1}^{m} (\mathbf{x}) := w^{k_1}_{n_1} (x_1) s^k_{n_2} (x_2) s^k_{n_3} (x_3) 
\label{A:29}
\eeq
\beq
w_{\mathbf{n},4,k_2,k_3}^{m} (\mathbf{x}) := s^k_{n_1} (x_1) w^{k2}_{n_2} (x_2) w^{k3}_{n_3} (x_3) 
\label{A:30}
\eeq
\beq
w_{\mathbf{n},5,k_1,k_2}^{m} (\mathbf{x}) := w^{k_1}_{n_1} (x_1) w^{k_2}_{n_2} (x_2) s^k_{n_3} (x_3) 
\label{A:31}
\eeq
\beq
w_{\mathbf{n},6,k_1,k_3}^{m} (\mathbf{x}) := w^{k_1}_{n_1} (x_1) s^k_{n_2} (x_2) w^{k_3}{n_3} (x_3) 
\label{A:32}
\eeq
\beq
w_{\mathbf{n},7,k_1,k_2,k_3}^{m} (\mathbf{x}) := w^{k_1}_{n_1} (x_1) w^{k_2}_{n_2} (x_2) w^{k_3}_{n_3} (x_3). 
\label{A:33}
\eeq
We call the functions $w_{\mathbf{n},\alpha}^{m} (\mathbf{x})$
generalized wavelets, where the index, $m$, represents the smallest
wavelet scale (largest $k_1,k_2,k_3$) appearing in the product.  The index
$\alpha$ indicates the values of $k_1,k_2,k_3$ as well as which of the
seven types of products appear in the basis function.

Since the scaling function is the solution of a renormalization group
equation and all of the other basis functions are linear combinations
of translated and scale-transformed scaling functions, it follows that
all of the basis functions have a fractal structure.  This makes it
difficult to represent the basis functions in terms of elementary
functions that are smooth on a sufficiently small scale.  Fortunately,
while it is possible to compute the scaling function to any desired
accuracy, this is never needed in any application.  What is needed are
the overlap integrals of products of arbitrary numbers of basis
functions and their derivatives.  It turns out that these integrals
satisfy renormalization group equations and can all be computed
exactly using the renormalization group equations and the
normalization condition, without knowing the value of the integrand at
any points.  These equations can also be used to compute the integrals
of products of these functions with polynomials of arbitrary degree.
Since the basis functions have compact support, and any continuous
function can be approximated by a polynomial on a compact interval, it
follows that integrals of products of these basis functions and
continuous functions can be computed to any desired accuracy.  Methods
for computing these quantities are discussed in section five and the appendix.

\section{Free field conventions}

In this section we summarize our conventions for a free mass-$\mu$
scalar field and list some standard results that will be used in the
subsequent sections.  Since the free field operators exist and the
wavelet basis expansions are exact, free fields provide a useful
laboratory for understanding truncations as approximations.

Free fields also play a role in modeling interacting theories.  In the
presence of both resolution and volume cutoffs, which are natural in
the wavelet basis, all of the operator products in the Hamiltonain
become finite sums of well-defined operators on the free-field Fock space, 
allowing one to formally work in the interaction representation.  
The renormalization group, discussed in section seven, provides one 
means to investigate the existence of a limiting interacting theory 
with an infinite number of degrees of freedom. 

The Lagrangian density for a free scalar field of mass $\mu$ is
\beq
{\cal L}(x) := 
{1 \over 2}(\dot{\Phi}(x)\dot{\Phi}(x)
-\pmb{\nabla}\Phi (x) \cdot \pmb{\nabla}\Phi (x)
- \mu^2 \Phi(x)^2). 
\label{B:1}
\eeq
Lagrange's equations are
\beq
{\partial^2 \over \partial t^2}  \Phi(x)
- {\partial^2 \over \partial \mathbf{x}^2}  \Phi(x)
+ \mu^2 \Phi (x) = 0 . 
\label{B:2}
\eeq
The generalized momentum is
\beq
\Pi (x) = {\partial {\cal L} (x) \over \partial \dot\Phi }=
\dot{\Phi}(x)
\label{B:3}
\eeq
and the Hamiltonian is
\[
H= \int (\Pi(x) \dot{\Phi}(x) - {\cal L}(x) )d\mathbf{x} 
\]
\beq
={1 \over 2}\int ({\Pi}(x){\Pi}(x)
+\pmb{\nabla}\Phi (x) \cdot \pmb{\nabla}\Phi (x)
+ \mu^2 \Phi(x)^2)d\mathbf{x} .
\label{B:4}
\eeq

The field $\Phi(\mathbf{x},t)$ is normalized so $\Phi (\mathbf{x},t)$
and $\Pi (\mathbf{x},t)$ satisfy the canonical equal-time commutation
relations
\beq
[\Phi (\mathbf{x},t), \Pi (\mathbf{y},t)]=i \delta (\mathbf{x}-\mathbf{y}).
\label{B:5}
\eeq

The classical energy-momentum tensor is
\[
T^{\mu\nu}(x) = - {\partial {\cal L}(x) \over \partial (\partial_{\mu} \Phi)}
\partial^{\nu}  \Phi + \eta^{\mu \nu} {\cal L}(x) =
\]
\beq
\partial^{\mu}\Phi (x) \partial^{\nu} \Phi(x) -
{1 \over 2}\eta^{\mu \nu}(\partial^\alpha \Phi(x) \partial_{\alpha} \Phi(x) 
+ \mu^2 \Phi(x)^2) 
\label{B:6}
\eeq
and the angular momentum tensor is 
\beq
M^{\mu \nu\alpha } = (x^{\mu} T^{\nu \alpha } - x^{\nu} T^{\mu \alpha } ). 
\label{B:7}
\eeq
These tensors satisfy the conservation laws 
\beq
\partial_{\mu}T^{\nu \mu} = \partial_\alpha M^{ \mu \nu \alpha }=0
\label{B:8}
\eeq
leading to the conserved charges, which after quantization, are the
infinitesimal generators of the Poincar\'e group
\beq
P^{\mu} = \int_{t=0} d\mathbf{x} {\cal P}^{\mu} (x) d\mathbf{x} 
\qquad
{\cal P}^{\mu}(x) := T^{\mu 0} (x)
\label{B:9}
\eeq
\beq
J^{\mu \nu} := \int_{t=0} d\mathbf{x} {\cal J}^{\mu\nu} (x) d\mathbf{x}
\qquad
{\cal J}^{\mu\nu }(x) := M^{\mu\nu 0} (x).
\label{B:10}
\eeq
The generators can be expressed as integrals 
over operator-valued densities
\[
H= P^0 = 
\]
\beq
\int_{t=0}: (\Pi(x) \dot{\Phi}(x) - {\cal L}(x) ):d\mathbf{x} =
{1 \over 2}\int : ({\Pi}(x){\Pi}(x)
+\pmb{\nabla}\Phi (x) \cdot \pmb{\nabla}\Phi (x)
+ \mu^2 \Phi(x)^2):d\mathbf{x}
\label{B:11}
\eeq
\beq
\mathbf{P} = -\int_{t=0} : \Pi(x) \pmb{\nabla} \Phi (x): d\mathbf{x} 
\label{B:12}
\eeq
\[
J^{ij}=-J^{ji} = \epsilon^{ijk} J^k =
\]
\beq
\int_{t=0} d\mathbf{x} :(\Pi (x) x^j \partial^i \Phi (x) 
-\Pi (x) x^i \partial^j \Phi (x) ):
\label{B:13}
\eeq
\[
J^{i0}=-J^{0i} = K^i = 
\]
\beq
\int_{t=0} :\left ( 
{1 \over 2} \Pi^2(x) x^i + {1 \over 2} x^i 
\pmb{\nabla} \Phi(x)\cdot \pmb{\nabla} \Phi(x) 
+ {1 \over 2} x^i \mu^2 \Phi (x)^2 
\right ): d\mathbf{x}
\label{B:14}
\eeq
where the $:$s indicate that the operators are normal
ordered.

The field has the Fourier representation ($-+++$ metric) 
\beq
\Phi (x) = {1 \over (2 \pi)^{3/2}}\int 
{d \mathbf{p} \over \sqrt{2 \omega_\mu (\mathbf{p}) }}
( e^{-i p\cdot x} a^{\dagger}(\mathbf{p})+  e^{i p\cdot x} a(\mathbf{p}))
\label{B:15}
\eeq
\beq
\Pi (x) = {i \over (2 \pi)^{3/2}}\int 
d \mathbf{p}\sqrt{{\omega_\mu (\mathbf{p}) \over 2}}
( e^{-i p\cdot x} a^{\dagger}(\mathbf{p})-  e^{i p\cdot x} a(\mathbf{p}))
\label{B:16}
\eeq
where 
\beq
\omega_\mu(\mathbf{p}) = \sqrt{\mu^2 + \mathbf{p}^2}
\eeq
is the single-particle energy.  The equal-time commutation relations
imply
\beq
[a (\mathbf{p}),a^{\dagger} (\mathbf{p}') ] = \delta (\mathbf{p}- \mathbf{p}') .
\label{B:17}
\eeq
The vacuum state of the field is the solution of 
\beq
a(\mathbf{p}) \vert 0 \rangle = 0 .
\label{B:18}
\eeq
The infinitesimal generators have momentum-space representations as
integrals over momentum densities
\beq
H = \int {\cal H}(\mathbf{p}) d\mathbf{p} = \int d \mathbf{p}  a^{\dagger} 
(\mathbf{p}) \omega_m (\mathbf{p})
a (\mathbf{p}), 
\label{B:19}
\eeq
\beq
\mathbf{P} = \int \pmb{{\cal P}}(\mathbf{p}) d\mathbf{p} = \int d \mathbf{p}  a^{\dagger} (\mathbf{p}) \mathbf{p}
a (\mathbf{p}), 
\label{B:20}
\eeq
\beq
\mathbf{J} = \int \pmb{{\cal J}}(\mathbf{p}) d\mathbf{p} = \int d \mathbf{p}  a^{\dagger} (\mathbf{p})
(i {\partial \over \partial \mathbf{p} } \times \mathbf{p} ) 
a(\mathbf{p}), 
\label{B:21}
\eeq
\beq
\mathbf{K} = \int \pmb{{\cal K}}(\mathbf{p}) d\mathbf{p} = \int d \mathbf{p}  a^{\dagger} (\mathbf{p})
{1 \over 2} \{ i {\partial \over \partial \mathbf{p} },
 \omega_m (\mathbf{p}) \} 
a(\mathbf{p}) 
\label{B:22}
\eeq
where we set $x^0=0$ in the last expression. 

\section{Wavelet discretized fields}

The wavelet basis discussed in (\ref{A:26}-\ref{A:33}) can be used to
write {\it exact} expansions of the fields as sums of products of
basis functions with time-dependent operator coefficients,
\beq
\Phi (\mathbf{x},t) =
\sum_{\mathbf{n}} \Phi^k (\mathbf{n},t) s_{\mathbf{n}}^{k} (\mathbf{x}) 
+ \sum_{\mathbf{n},\alpha,l \geq k} 
\Phi^l (\mathbf{n},\alpha,t) w^l_{\mathbf{n}\alpha } (\mathbf{x}) 
\label{C:1}
\eeq
\beq
\Pi (\mathbf{x},t) =
\sum_{\mathbf{n}} \Pi^k (\mathbf{n},t) s_{\mathbf{n}}^{k} (\mathbf{x}) 
+ \sum_{\mathbf{n},\alpha,l \geq k} 
\Pi^l (\mathbf{n},\alpha,t) w^l_{\mathbf{n}\alpha } (\mathbf{x}), 
\label{C:2}
\eeq
where the operator coefficients are projections of the field operators
on the orthonormal basis of scaling functions and wavelets
\beq
\Phi^k (\mathbf{n},t) = \int d\mathbf{x} 
\Phi (\mathbf{x},t) s_{\mathbf{n}}^{k} (\mathbf{x})
\label{C:3}
\eeq
\beq
\Phi^l (\mathbf{n},\alpha,t) = 
\int d\mathbf{x} 
\Phi (\mathbf{x},t)
w^l_{\mathbf{n}\alpha } (\mathbf{x}) \qquad (l \geq k)
\label{C:4}
\eeq
\beq
\Pi^k (\mathbf{n},t) = \int d\mathbf{x} 
\Pi (\mathbf{x},t) s_{\mathbf{n}}^{k} (\mathbf{x})
\label{C:5}
\eeq
\beq
\Pi^l (\mathbf{n},\alpha,t) = 
\int d\mathbf{x} 
\Pi (\mathbf{x},t)
w^l_{\mathbf{n}\alpha } (\mathbf{x}) \qquad (l \geq k ) . 
\label{C:6}
\eeq
In these expressions the spacetime points have dimensions.  If $L$ is
a length scale then replacing $s^k_m(x)$ and $w^l_n(x)$ by ${1
\over \sqrt{L}} s^k_m({x\over L})$ and ${1 \over \sqrt{L}}
w^l_n({x\over L})$ preserves all of the formulas in the previous
section, except the translations and rescaling are in units of $L$.
The scale factors $1/2^k$ remain dimensionless.  All factors of $L$
disappear if we make this change and use natural units where $L=1$.

In these expression we only integrate over the spatial
coordinates $\mathbf{x}$ so the operator coefficients are time
dependent.  It is also possible to integrate over all space-time
coordinates, which is useful in action-based formulations.

These expansions have been separated into operators smeared over scaling 
functions and generalized wavelets.  This is a separation of scales - the 
scale-$1/2^k$ scaling functions give the coarse-scale structure of the field,
while the wavelet parts include the structure on all smaller scales.

The field operators smeared with the scaling functions represent the
average value of the field over a compact region weighted by the
product of scaling functions.  Because the scale $1/2^{k+1}$ scaling
function can be expressed in terms of the scale $1/2^k$ scaling
functions and scale $1/2^k$ wavelets, (\ref{A:18}), the scale $1/2^k$
wavelet and scaling-function smeared fields, contain the same
information as the scale $1/2^{k+1}$ scaling-function smeared fields.

When all of the wavelet contributions are included the expansions are exact.
In any finite region there are still an infinite number of wavelet
basis functions associated with arbitrarily small scales.

For fields satisfying canonical equal-time commutation relations, 
the orthonormality of the basis functions implies that the equal-time
canonical commutation relations for the discrete fields are:
\beq
[\Phi^k (\mathbf{n},t),\Phi^k (\mathbf{m},t)]=0
\qquad
[\Pi^k (\mathbf{n},t),\Pi^k (\mathbf{m},t)]=0
\label{C:7}
\eeq
\beq
[\Phi^k (\mathbf{n},t),\Pi^k (\mathbf{m},t)]=i \delta_{\mathbf{n},\mathbf{m}}
\label{C:8}
\eeq
\beq
[\Phi^r (\mathbf{n},\alpha, t),\Phi^s (\mathbf{m},\beta, t)]=0
\qquad
[\Pi^r (\mathbf{n},\alpha,t),\Pi^s (\mathbf{m},\beta, t)]=0
\label{C:9}
\eeq
\beq
[\Phi^r (\mathbf{n},\alpha,t),\Pi^s (\mathbf{m},\beta,t)]=i 
\delta_{\alpha \beta} \delta_{rs} \delta_{\mathbf{n},\mathbf{m}}
\label{C:10}
\eeq
\beq
[\Phi^r (\mathbf{n},\alpha, t),\Phi^s (\mathbf{m}, t)]=0
\qquad
[\Pi^r (\mathbf{n},\alpha, t),\Pi^s (\mathbf{m}, t)]=0
\label{C:11}
\eeq
\beq
[\Phi^r (\mathbf{n},\alpha, t),\Pi^s (\mathbf{m}, t)]=0
\qquad
[\Pi^r (\mathbf{n},\alpha, t),\Phi^s (\mathbf{m}, t)]=0
\label{C:12}
\eeq
where in all of these expressions $k\leq r,s$.

Given these discrete field operators we can construct
discrete creation and annihilation operators
\beq
a^k(\mathbf{n},t) : = {1 \over \sqrt{2} } 
(\sqrt{\gamma} \Phi^k (\mathbf{n},t) +i {1 \over \sqrt{\gamma}} 
\Pi^k (\mathbf{n},t))
\label{C:13}
\eeq
\beq
b^r(\mathbf{n},\alpha, t) : ={1 \over \sqrt{2} }((\sqrt{\gamma}  
\Phi^k (\mathbf{n},\alpha,t)+i {1 \over \sqrt{\gamma}} \Pi^k (\mathbf{n},\alpha ,t))
\label{C:14}
\eeq
that satisfy 
\beq
[a^k(\mathbf{n},t),a^{k\dagger}(\mathbf{m},t)] = \delta_{\mathbf{m}\mathbf{n}}
\label{C:15}
\eeq
\beq 
[b^l(\mathbf{n},\alpha,t),b^{j\dagger}(\mathbf{m},\beta, t)] = 
\delta_{\mathbf{m}\mathbf{n}}\delta_{jl}\delta_{\alpha \beta} 
\label{C:16}
\eeq
with all other commutators vanishing.  In these expressions $\gamma$
is a constant that can depend on the discrete indices.  While these
quantities and their adjoints satisfy the commutation relations
({\ref{C:15},\ref{C:16}) for creation and annihilation operators for
any choice of $\gamma$, a specific choice is required if the
annihilation operator is to annihilate the free field vacuum.  This
requires that $\gamma$ is chosen so
\beq
\langle 0 \vert a^{k\dagger}  (\mathbf{n},t)
a^{k} (\mathbf{n},t) \vert 0 \rangle =0 
\label{C:17}
\eeq
\beq
\langle 0 \vert b^{r\dagger}  (\mathbf{n},\alpha ,t)
b^{r} (\mathbf{n},\alpha ,t) \vert 0 \rangle =0 
\label{C:18}
\eeq 
\beq
\langle 0 \vert  a^{k} (\mathbf{n},t)   a^{k\dagger} 
(\mathbf{n},t) \vert 0 \rangle =1 
\label{C:19}
\eeq
\beq
\langle 0 \vert 
b^{r}  (\mathbf{n},\alpha ,t) b^{r\dagger}  (\mathbf{n},\alpha ,t) 
\vert 0 \rangle =1 
\label{C:20}
\eeq 
which leads to quadratic equations for the coefficients $\gamma$.
With this choice of $\gamma$ the annihilation operators annihilate the 
mass $\mu$ free-field vacuum.

The solution of these equations for the scaling-function fields is 
\[
\gamma (k)= { 1 \pm \sqrt{1 - 4 \langle 0 \vert \Phi^k (\mathbf{m},t) 
\Phi^k (\mathbf{m},t)\vert 0 \rangle
\langle 0 \vert \Pi^k (\mathbf{m},t)\Pi^k (\mathbf{m},t) \vert 0 \rangle} \over 
2 \langle 0 \vert \Phi^k (\mathbf{m},t)\Phi^k (\mathbf{m},t) \vert 0 \rangle}  
\]
\beq
={ 1 \pm \sqrt{1 - 4 \langle 0 \vert \Phi^k (\mathbf{0},0) 
\Phi^k (\mathbf{0},0)\vert 0 \rangle
\langle 0 \vert \Pi^k (\mathbf{0},0)\Pi^k (\mathbf{0},0) \vert 0 \rangle} \over 
2 \langle 0 \vert \Phi^k (\mathbf{0},0)\Phi^k (\mathbf{0},0) \vert 0 \rangle} 
\label{C:21}
\eeq
and for the wavelet fields
\[
\gamma (k,\alpha )= { 1 \pm \sqrt{1 - 4 \langle 0 \vert \Phi^r
    (\mathbf{m},\alpha,t) \Phi^r (\mathbf{m},\alpha,t)\vert 0 \rangle
    \langle 0 \vert \Pi^r (\mathbf{m},\alpha,t)\Pi^r
    (\mathbf{m},\alpha,t) \vert 0 \rangle} \over 2 \langle 0 \vert
  \Phi^r (\mathbf{m},\alpha,t)\Phi^r (\mathbf{m},\alpha,t) \vert 0
  \rangle} 
\]
\beq
={ 1 \pm \sqrt{1 - 4 \langle 0 \vert \Phi^r
    (\mathbf{0},\alpha,0) \Phi^r (\mathbf{0},\alpha,0)\vert 0 \rangle
    \langle 0 \vert \Pi^r (\mathbf{0},\alpha,0)\Pi^r
    (\mathbf{0},\alpha,0) \vert 0 \rangle} \over 2 \langle 0 \vert
  \Phi^r (\mathbf{0},\alpha,0)\Phi^r (\mathbf{0},\alpha,0) \vert 0
  \rangle}.
\label{C:22}
\eeq
These coefficients depend on the scale $1/2^k$, $\alpha$ and the mass
$\mu$; however they are independent of $\mathbf{m}$ and $t$ by the
space-time translational invariance of the vacuum.  With this choice
of $\gamma$ the scaling function fields can be expressed in terms of
the creation and annihilation operators as
\beq
\Phi^k (\mathbf{m},t) = {1 \over \sqrt{ 2 \gamma}} 
(a^{k\dagger}(\mathbf{m},t) + a^k(\mathbf{m},t))
\label{C:23}
\eeq
\beq
\Pi^k (\mathbf{m},t) = i \sqrt{{\gamma \over 2 }} 
(a^{k\dagger}(\mathbf{m},t) - a^k(\mathbf{m},t))
\label{C:24}
\eeq
with analogous expressions for the wavelet fields,
\beq
\Phi^k (\mathbf{m},\alpha,t) = {1 \over \sqrt{ 2 \gamma}} 
(b^{k\dagger}(\mathbf{m},\alpha,t) + b^k(\mathbf{m},\alpha,t))
\label{C:25a}
\eeq
\beq
\Pi^k (\mathbf{m},\alpha,t) = i \sqrt{{\gamma \over 2}} 
(b^{k\dagger}(\mathbf{m},\alpha,t) - b^k(\mathbf{m},\alpha, t)).
\label{C:26a}
\eeq
The coefficients $\gamma$ depend on the mass term in the field operators. This
is because the following integrals appear in (\ref{C:21}-\ref{C:22}):
\beq
\langle 0 \vert \Phi^k (\mathbf{m},t) 
\Phi^k (\mathbf{m},t)\vert 0 \rangle =
{1 \over (2 \pi)^3} \int {s_{\mathbf{n}}^k(\mathbf{x}) s_{\mathbf{n}}^k(\mathbf{y}) \over 2 
\omega_\mu(\mathbf{p})} e^{i \mathbf{p} \cdot (\mathbf{x}-\mathbf{y})}
d\mathbf{x} d\mathbf{y} d\mathbf{p}  
\label{C:25}
\eeq
\beq
\langle 0 \vert \Pi^k (\mathbf{n},t) 
\Pi^k (\mathbf{n},t)\vert 0 \rangle =
{1 \over (2 \pi)^3} \int {s_{\mathbf{n}}^k(\mathbf{x}) s_{\mathbf{n}}^k
(\mathbf{y}) 
\omega_\mu(\mathbf{p}) \over 2 
} e^{i \mathbf{p} \cdot (\mathbf{x}-\mathbf{y})} 
d\mathbf{x} d\mathbf{y} d\mathbf{p} 
\label{C:26}
\eeq
\beq
\langle 0 \vert \Phi^r (\mathbf{m},\alpha,t) 
\Phi^r (\mathbf{m},\alpha,t)\vert 0 \rangle =
{1 \over (2 \pi)^3} \int {w_{\alpha,\mathbf{n}}^r(\mathbf{x}) 
w_{\alpha,\mathbf{n}}^r(\mathbf{y}) \over 2 
\omega_\mu(\mathbf{p})} e^{i \mathbf{p} \cdot (\mathbf{x}-\mathbf{y})}  
d\mathbf{x} d\mathbf{y} d\mathbf{p}
\label{C:27}
\eeq
\beq
\langle 0 \vert \Pi^r (\mathbf{n},\alpha,t) 
\Pi^r (\mathbf{n},\alpha,t)\vert 0 \rangle =
{1 \over (2 \pi)^3} \int {w_{\alpha,\mathbf{n}}^r(\mathbf{x}) 
w_{\alpha,\mathbf{n}}^r(\mathbf{x}) 
\omega_\mu(\mathbf{p}) \over 2 
} e^{i \mathbf{p} \cdot (\mathbf{x}-\mathbf{y})}
d\mathbf{x} d\mathbf{y} d\mathbf{p} . 
\label{C:28}
\eeq
These are integrals of the basis functions over the two-point
mass-$\mu$ Wightman functions of this field at fixed time.  
 

Using (\ref{C:23}-\ref{C:28}) in (\ref{C:1}-\ref{C:2}) gives 
exact expressions for the field operators in terms of the 
discrete creation and annihilation operators:
\[
\Phi (x) = \sum {1 \over \sqrt{2 \gamma (k)}} s^k_{\mathbf{m}}(x)
(a^k(\mathbf{m},t) + a^{k\dagger} (\mathbf{m},t)) 
\]
\beq 
+ \sum  {1 \over \sqrt{2 \gamma (j,\alpha)}} w^j_{\mathbf{m},\alpha}(x)
(b^j(\mathbf{m},\alpha,t) + b^{j\dagger} (\mathbf{m},\alpha, t)) 
\label{C:29}
\eeq
\[
\Pi (x) = i\sum  {\sqrt{\gamma (k) \over 2 }} s^k_{\mathbf{m}}(x)
(a^{k\dagger} (\mathbf{m},t) - a^k(\mathbf{m},t))   
\]
\beq 
+ i \sum  {\sqrt{\gamma (j,\alpha)  \over 2}} w^j_{\mathbf{m},\alpha}(x)
(b^{j\dagger} (\mathbf{m},\alpha, t) - b^j(\mathbf{m},\alpha,t) ) .
\label{C:30}
\eeq
In this notation the $a$ and $a^{\dagger}$ operators destroy and create 
scale $1/2^k$ degrees of freedom, while the $b$ and $b^{\dagger}$ operators
destroy and create smaller-scale degrees of freedom.

The Hilbert space for this free field is generated by taking
limits of finite linear combinations of products
of the discrete creation operators,
$a^k(\mathbf{m},t)^{\dagger},b^l(\mathbf{m},\alpha,t)^{\dagger}$
applied to the vacuum at a fixed time.

It is useful to decompose operators into parts that only involve the 
$a$ and $a^{\dagger}$ operators,  parts that only involve the 
$b$ and $b^{\dagger}$ operators, and mixed terms involving products of
at least one operator from each of the above groups.

The terms with only the $a$ and $a^{\dagger}$ operators represent the
$2^{-k}$-scale physics, the terms with only the $b$ and
$b^{\dagger}$ operators represent the part of the operator associated with
scales finer than $2^{-k}$ that do not couple to the scale $2^{-k}$
operators, and the mixed terms are responsible for the coupling of the
$2^{-k}$-scale degrees of freedom to the smaller-scale degrees of
freedom.

The Hamiltonian (\ref{B:11}) has the decomposition
\beq
H= H(a) + H(b) + H(ab) 
\label{C:31}
\eeq
where 
\[
H(a) :=  
{1 \over 2} (\sum_{\mathbf{n}} :{\Pi}^k(\mathbf{n},0){\Pi}^k(\mathbf{n},0):
+\sum_{\mathbf{m}\mathbf{n}}:\Phi^k (\mathbf{m} ,0) D^k_{\mathbf{m}\mathbf{n}}
\Phi^k (\mathbf{n},0):
\]
\beq
+ \mu^2 \sum_{\mathbf{n}} :{\Phi}^k(\mathbf{n},0){\Phi}^k(\mathbf{n},0):)
\label{C:32}
\eeq
\[
H(b) :=  
{1 \over 2} (\sum_{\mathbf{n}\alpha,l} :{\Pi}^{l}(\mathbf{n},\alpha,0)
{\Pi}^l(\mathbf{n},\alpha,0):
\]
\[
+ \sum_{\mathbf{m}\alpha,l\mathbf{n}\beta,j}:\Phi^l (\mathbf{m},\alpha ,0) 
D^{lj}_{\mathbf{m},\alpha,\mathbf{n},\beta}
\Phi^j (\mathbf{n},\beta,0):
\]
\beq
+ \mu^2 \sum_{l,\mathbf{n},\alpha} :{\Phi}^l(\mathbf{n},\alpha,0)
{\Phi}^l(\mathbf{n},\alpha,0):)
\label{C:33}
\eeq
\beq
H(ab) :=  
{1 \over 2} 
\sum_{\mathbf{m}\alpha,l\mathbf{n}}:\Phi^l (\mathbf{m},\alpha ,0) 
D^{lk}_{\mathbf{m},\alpha,\mathbf{n}}
\Phi^k (\mathbf{n},0):
\label{C:34}
\eeq
and the coefficients 
$D^k_{\mathbf{m}\mathbf{n}}$,
$D^{lj}_{\mathbf{m},\alpha,\mathbf{n},\beta}$ and 
$D^{lk}_{\mathbf{m},\alpha,\mathbf{n}}$ are 
given by 
\beq
D^k_{\mathbf{m}\mathbf{n}}  = \int d \mathbf{x} 
\pmb{\nabla} s^k_{\mathbf{m}} (\mathbf{x}) \cdot 
\pmb{\nabla} s^k_{\mathbf{n}} (\mathbf{x})
\label{C:35}
\eeq
\beq
D^{lj}_{\mathbf{m},\alpha,\mathbf{n},\beta}= 
\int d \mathbf{x} 
\pmb{\nabla} w^l_{\mathbf{m},\alpha} (\mathbf{x}) \cdot 
\pmb{\nabla} w^j_{\mathbf{n},\beta} (\mathbf{x})
\label{C:36}
\eeq
\beq
D^{lk}_{\mathbf{m},\alpha,\mathbf{n}}=
2 \int d \mathbf{x} 
\pmb{\nabla} w^l_{\mathbf{m},\alpha} (\mathbf{x}) \cdot 
\pmb{\nabla} s^k_{\mathbf{n}} (\mathbf{x}) .
\label{C:37}
\eeq
The derivatives of the basis functions in (\ref{C:35}-\ref{C:37})
exist for Daubechies $K\geq 3$ scaling functions.  The computation of
these integrals, using the renormalization group equations and
normalization condition, is discussed in the next section and the
appendix.  These integrals are almost local in the sense that they
vanish when the support of the basis functions do not overlap.  From
the above equations we see that the part of the Hamiltonian, $H(ab)$,
that couples the scale $1/2^k$ degrees of freedom to the finer scale 
degrees of freedom proceeds through the 
terms in the free-field Hamiltonian with the spatial derivatives.
We will see that interactions also lead to coupling of scales.

There are similar decompositions for the linear momentum (\ref{B:12}),
angular momentum (\ref{B:13}), and Lorentz boost generators (\ref{B:14}).  
For the linear momentum the decomposition is:
\beq
\mathbf{P} = \mathbf{P}(a)+ \mathbf{P}(b) + \mathbf{P}(ab)
\label{C:38}
\eeq
where 
\beq
\mathbf{P}(a) = - \sum_{\mathbf{m}\mathbf{n}}
:\Pi^k (\mathbf{m} ,0) \mathbf{P}^k_{\mathbf{m}\mathbf{n}}
\Phi^k (\mathbf{n},0):
\label{C:39}
\eeq
\beq
\mathbf{P}(b) = 
- \sum_{\mathbf{m}\alpha,l\mathbf{n}\beta,j}:\Pi^l (\mathbf{m},\alpha ,0) 
\mathbf{P}^{lj}_{\mathbf{m},\alpha,\mathbf{n},\beta}
\Phi^j (\mathbf{n},\beta,0):
\label{C:40}
\eeq
\beq
\mathbf{P}(ab)=
- \sum_{\mathbf{m}\alpha,l\mathbf{n}}:\Pi^l (\mathbf{m},\alpha ,0) 
\mathbf{P}^{lk}_{\mathbf{m},\alpha,\mathbf{n}}
\Phi^k (\mathbf{n},0):,
\label{C:41}
\eeq
and
\beq
\mathbf{P}^{k}_{\mathbf{m},\mathbf{n}}
=
\int d \mathbf{x} 
 s^k_{\mathbf{m}} (\mathbf{x})  \pmb{\nabla}_{\mathbf{x}}
s^k_{\mathbf{n}} (\mathbf{x})
\label{C:42}
\eeq
\beq
\mathbf{P}^{lj}_{\mathbf{m},\alpha,\mathbf{n},\beta}
=
\int d \mathbf{x} 
 w^l_{\mathbf{m},\alpha} (\mathbf{x})  \pmb{\nabla}_{\mathbf{x}}
w^j_{\mathbf{n},\beta} (\mathbf{x}) + 
\label{C:43}
\eeq
\beq
\mathbf{P}^{l}_{\mathbf{m},\alpha,\mathbf{n}}
=
\int d \mathbf{x} 
(w^l_{\mathbf{m},\alpha} (\mathbf{x})  \pmb{\nabla}_{\mathbf{x}}
s^k_{\mathbf{n}} (\mathbf{x}) + 
s^k_{\mathbf{n}} (\mathbf{x})
\pmb{\nabla}_{\mathbf{x}}  w^l_{\mathbf{m},\alpha} (\mathbf{x}))  .
\label{C:44}
\eeq
For the angular momentum 
\beq
\mathbf{J} = \mathbf{J}(a)+ \mathbf{J}(b) + \mathbf{J}(ab)
\label{C:45}
\eeq
where 
\beq
\mathbf{J}(a) =  \sum_{\mathbf{m}\mathbf{n}}
:\Pi^k (\mathbf{m} ,0) \mathbf{J}^k_{\mathbf{m}\mathbf{n}}
\Phi^k (\mathbf{n},0):
\label{C:46}
\eeq
\beq
\mathbf{J}(b) = 
\sum_{\mathbf{m}\alpha,l\mathbf{n}\beta,j}:\Pi^l (\mathbf{m},\alpha ,0) 
\mathbf{J}^{lj}_{\mathbf{m},\alpha,\mathbf{n},\beta}
\Phi^j (\mathbf{n},\beta,0):
\label{C:47}
\eeq
\beq
\mathbf{J}(ab)=
\sum_{\mathbf{m}\alpha,l\mathbf{n}}:
\Pi^l (\mathbf{m},\alpha ,0) \mathbf{J}^{lk}_{\mathbf{m},\alpha,\mathbf{n}}
\Phi^k (\mathbf{n},0): +
\Pi^k (\mathbf{n} ,0) \mathbf{J}^{kl}_{\mathbf{n},\mathbf{m},\alpha}
\Phi^l (\mathbf{m},\alpha ,0): 
\label{C:48}
\eeq
and
\beq
\mathbf{J}^{k}_{\mathbf{m},\mathbf{n}}
=
- \int d \mathbf{x} ( s^k_{\mathbf{m}} (\mathbf{x}) (\mathbf{x} \times  \pmb{\nabla}_{\mathbf{x}})
s^k_{\mathbf{n}} (\mathbf{x})
\label{C:49}
\eeq
\beq
\mathbf{J}^{lj}_{\mathbf{m},\alpha,\mathbf{n},\beta}
=
- \int d \mathbf{x} 
 w^l_{\mathbf{m},\alpha} (\mathbf{x}) (\mathbf{x} 
\times  \pmb{\nabla}_{\mathbf{x}})
w^j_{\mathbf{n},\beta} (\mathbf{x}), 
\label{C:50}
\eeq
\beq
\mathbf{J}^{lk}_{\mathbf{m},\alpha,\mathbf{n}}
=
- \int d \mathbf{x} 
 w^l_{\mathbf{m},\alpha} (\mathbf{x}) (\mathbf{x} \times  
\pmb{\nabla}_{\mathbf{x}})
s^k_{\mathbf{n}} (\mathbf{x}) ,
\eeq
\beq
\mathbf{J}^{kl}_{\mathbf{m},\alpha,\mathbf{n}} =
- \int d \mathbf{x} 
s^k_{\mathbf{n}} (\mathbf{x})(\mathbf{x} \times  
\pmb{\nabla}_{\mathbf{x}})  w^l_{\mathbf{m},\alpha} (\mathbf{x})  .
\label{C:51}
\eeq
The decomposition for the rotationless boost generators is 
\beq
\mathbf{K} = \mathbf{K}(a)+ \mathbf{K}(b) + \mathbf{K}(ab).
\label{C:52}
\eeq
where 
\[
\mathbf{K}(a) = 
\int 
{1 \over 2} \sum_{\mathbf{m},\mathbf{n}}
\left (
\mathbf{F}^k_{\mathbf{n}\mathbf{m}} (
:{\Pi}(\mathbf{n},0) \Pi(\mathbf{m},0):
\right .
\]
\beq
\left .
+ \mu^2 :{\Phi}(\mathbf{n},0) \Phi(\mathbf{m},0):) + 
\mathbf{G}_{\mathbf{n}\mathbf{m}}
 :{\Phi}(\mathbf{n},0) \Phi(\mathbf{m},0): 
\right )
\label{C:53}
\eeq
\[
\mathbf{K}(b) = 
\int 
{1 \over 2} \sum_{l\mathbf{m}\alpha ,j\mathbf{n}\beta}
\left (
\mathbf{F}^{lj}_{\mathbf{m},\alpha ,\mathbf{n},\beta} (
:{\Pi}^l(\mathbf{m},\alpha) \Pi^j(\mathbf{n},\beta,0):
\right .
\]
\beq
\left .
+ \mu^2 :{\Phi}^l(\mathbf{m},\alpha,0) \Phi^j(\mathbf{n},\beta,0):) + 
\mathbf{G}_{\mathbf{m}\alpha \mathbf{n}\beta}^{lj}
 :{\Phi}^l(\mathbf{m},\alpha,0) \Phi^j(\mathbf{n},\beta,0): 
\right )
\label{C:54}
\eeq
\[
\mathbf{K}(ab) = 
\int 
\sum_{l,\mathbf{m},\alpha ,\mathbf{n}}
\left (
\mathbf{F}^{lk}_{\mathbf{m}\alpha ,\mathbf{n}} (
:{\Pi}^l(\mathbf{m},\alpha) \Pi^k(\mathbf{n},0):
\right .
\]
\beq
\left .
+ \mu^2 :{\Phi}^l(\mathbf{m},\alpha,0) \Phi^k(\mathbf{n},0):) + 
\mathbf{G}^{lk}_{\mathbf{n},\alpha,\mathbf{m}}
:{\Phi}^l(\mathbf{n},\alpha,0) \Phi^k(\mathbf{m},0): 
\right )
\label{C:55}
\eeq
and the overlap integrals are
\beq
\mathbf{F}^k_{\mathbf{n}\mathbf{m}} = \int d \mathbf{x} 
s^k_{\mathbf{n}} (\mathbf{x}) \mathbf{x} 
s^k_{\mathbf{m}} (\mathbf{x})
\label{C:56}
\eeq
\beq
\mathbf{G}^k_{\mathbf{n}\mathbf{m}} = \int d \mathbf{x} 
\pmb{\nabla} s^k_{\mathbf{n}} (\mathbf{x}) \cdot 
\pmb{\nabla} s^k_{\mathbf{m}} (\mathbf{x}) \mathbf{x}
\label{C:57}
\eeq
\beq
\mathbf{F}^{lj}_{\mathbf{n}\alpha\mathbf{m}\beta} = \int d \mathbf{x} 
w^l_{\mathbf{n},\alpha} (\mathbf{x}) \mathbf{x} 
w^j_{\mathbf{m},\beta} (\mathbf{x})
\label{C:58}
\eeq
\beq
\mathbf{G}^{lj}_{\mathbf{n}\alpha\mathbf{m}\beta} = \int d \mathbf{x} 
\pmb{\nabla} w^l_{\mathbf{n}\alpha} (\mathbf{x}) \cdot 
\pmb{\nabla} w^j_{\mathbf{m}\beta} (\mathbf{x}) \mathbf{x}
\label{C:59}
\eeq
\beq
\mathbf{F}^{lk}_{\mathbf{n}\alpha\mathbf{m}} = \int d \mathbf{x} 
w^l_{\mathbf{n},\alpha} (\mathbf{x}) \mathbf{x} 
s^k_{\mathbf{m}} (\mathbf{x})
\label{C:60}
\eeq
\beq
\mathbf{G}^{lk}_{\mathbf{n}\alpha\mathbf{m}} = \int d \mathbf{x} 
\pmb{\nabla} w^l_{\mathbf{n}\alpha} (\mathbf{x}) \cdot 
\pmb{\nabla} s^k_{\mathbf{m}} (\mathbf{x}) \mathbf{x}.
\label{C:61}
\eeq
All of the numerical coefficients can be computed exactly using
linear algebra and the scaling equation.  They vanish when the
supports of the wavelets or scaling functions have no overlap.
In addition, it is a simple matter to determine how they scale
with $1/2^k$.

\section{Integrals}

The expressions for the Poincar\'e generators derived in the previous
section are linear combinations of normal products of discrete wavelet
fields multiplied by numerical coefficients.  The numerical
coefficients are integrals of products of scaling functions, wavelets,
first derivatives of these quantities and powers of $\mathbf{x}$.  For
a $3+1$ dimensional field theory each coefficient is a product of
three coefficients involving one-dimensional integrals.  In this
section we discuss methods for computing these one-dimensional
integrals.

While all of the integrals involve functions with fractal values,
making standard numerical methods impractical, the unique properties
of the wavelet basis makes it possible to compute all of the integrals
appearing in the previous section exactly.  In this section we discuss
how to compute the coefficients in the previous section.  We follow
the general discussion with an illustrative example.

The integrals in the last section are products of one-dimensional 
integrals of the form 
\beq
\Gamma_{n} 
= \int f_{1}({x}) f_{2}
({x}) \cdots f_{n} ({x}) 
d{x} .
\label{D:1}
\eeq
where the functions $f_{i}({x})$ are scaling functions,
wavelets, first derivatives of these functions or powers of $x$

The important observation is that all of the functions that appear in
the integrand of the coefficient $\Gamma_{n }$ have simple
transformation properties under scale transformation and translations.
This, along with the scaling equation and normalization condition, can
be used to reduce the computation of all of these quantities to finite
linear algebra.

The computation uses the following relations, which follow from 
the definitions:
\beq
\int s^k_n(x) dx = {1 \over \sqrt{2^k}} 
\label{D:2}
\eeq
\beq
D T^{2k} = T^k D
\label{D:3}
\eeq
\beq
{d \over dx} D = 2 D {d \over dx} 
\label{D:4}
\eeq
\beq 
Dx = 2 xD
\label{D:5} 
\eeq
\beq
Tx= (x-1)T .
\label{D:6} 
\eeq
In addition, we use the scaling equation, the definition of the wavelet
and the derivatives of these equations in the form (\ref{A:19}-\ref{A:22}):
\beq
s^k_m(x) = \sum_n H_{mn} s^{k+1}_{n} (x) 
\label{D:7}
\eeq
\beq
w^k_m(x) = \sum_n G_{mn} s^{k+1}_{n} (x) 
\label{D:8}
\eeq
\beq
s^{k\prime}_m = 2 \sum_n H_{mn} s^{(k+1)\prime}_{n}  
\label{D:9}
\eeq
\beq
w^{k\prime}_m = 2 \sum_n G_{mn} s^{(k+1)\prime}_{n} . 
\label{D:10}
\eeq
These equations can be used to express $s^k_m(x)$, $w^k_m(x)$,
$s^{k\prime}_m(x)$, and $w^{k\prime}_m(x)$, with scale $2^{-k}$, as
linear combinations of scaling functions and their derivatives with
scale $2^{-(k+1)}$.  Repeated application of these equations can be used
to increase $k$ in each of the functions by any desired amount.

In addition,  the scale factor $k$ of all functions in the integral 
can be increased or decreased by the same amount using
\[
\int D^k f_{1}(x) D^k f_{2}(x) \cdots D^k f_{m} (x)  dx =
\]
\beq
2^{k({m\over 2}-1)}
\int f_{1}(x) f_{2}(x) \cdots f_{m} (x)  dx 
\label{D:11}
\eeq
where in this equation $k$ can be positive or negative.  We recall
from the definition (\ref{A:4}) and (\ref{A:7}) that $D$ increases $k$
by 1 when applied to scaling functions or wavelets.  For derivatives
we also need to use (\ref{D:4}).

Using (\ref{D:11}) to make all of the $k$-values negative and then
using (\ref{D:7}-\ref{D:10}) to increase each one to zero leads to an
expression for the integral (\ref{D:1}) as a linear combination of
integrals involving products of scale $1/2^0$ scaling functions, 
their derivatives and powers of $x$.

It is useful to first consider integrals where there are no powers of
$x$;  we will see later that the corresponding integrals with powers of 
$x$ can be expressed in terms of the integrals with no powers of $x$.
To compute a general coefficient, $\Gamma_{n}$ we use the following steps.
\begin{itemize}

\item [1.] Step 1: Use (\ref{D:11}) to relate the integral to another integral
where the finest scale appearing in the integrand is $1/2^0$.  
Then the scale of each function in the integrand is $1/2^0$ or coarser
(negative $k$).  

\item [2.] Step 2: Use repeated application
(\ref{D:7}-\ref{D:10}) to replace all of the coarse scale functions by
linear combinations of scale $1/2^0$ functions.  The result of these
two steps is that original integral can be expressed as a finite sum
of coefficients where all of the functions in the integrand are scale
$1/2^0$ scaling functions or their derivatives.  Note that the wavelet
contributions can always be expressed in terms of scaling functions
using (\ref{D:8} or \ref{D:10}).

\item [3.] Step 3: Use integer translational 
invariance to make the support of the left most function start at 0.
Then, because each function has compact support on an interval of
width [2K-1], non-zero coefficients have translational indices that
vary from $[-2K+2,2K-2]$.  That means the number of non-zero
coefficients that cannot be generated by translations is less than
$(4K-3)^{n-1}$ (finite).

\item [4.] Step 4: Apply the scaling equation to the 
integrand of the resulting integral and use translational invariance
to shift the left-most index to zero.  This results in a set of
$(4K-3)^{n-1}$ homogeneous linear equations for the coefficient
functions with left-most index zero.

\item [5.] Step 5: Use the partition of unity property (there is also one for
the derivative of the scaling function) to get additional 
inhomogeneous equations for these coefficients.
  
\item [6.] Step 6: Combine the homogeneous and inhomogeneous equations to 
get a linear system and solve for the coefficients. 

\item [7.] Step 7: Reverse the steps,  using the finite number of 
solutions of the linear equations to construct the general coefficients.

\end{itemize} 

All of the steps are straightforward; for step six some of the equations are
redundant and one must select a set of independent equations that include 
at least one inhomogeneous equation.  

We illustrate construction with the example of computing the 
integral of the product of a scaling function, wavelet, and derivative 
of a scaling function with different scales
\beq
I_{mnp}^{k,k+l,k+j} 
:= \int s^k_{m} (x) w^{k+l}_n (x) {ds^{k+j}_p\over dx}  (x) dx .
\label{D:12}
\eeq
To be specific we assume that $j>l$ and $k>0$.  For the 
first step we use (\ref{D:11}) and (\ref{D:4}) 
to reduce the finest scale, $1/2^{k+j}$ to $1/2^0$   
\beq
I_{mnp}^{k,k+l,k+j}  = 
2^{-3(k+j)/2} I_{mnp}^{-j,-j+l,0}. 
\label{D:13}
\eeq
This makes the last scale index $0$ and the other two negative.  Since
all of the other scale factors have a negative exponent, we can use
(\ref{D:8}-\ref{D:9}) to increase each of them to zero:
\beq
I_{mnp}^{-j,-j+l,0} = \sum_{m'n'}  H^j_{mm'} (GH^{j-l-1})_{nn'}  
I_{m'n'p}^{0,0,0}
\label{D:14}
\eeq
where in (\ref{D:14}) $H^j_{mm'}$ is the product of $j$ factors of the
matrix $H_{mn}$.  While the matrix $H_{mn}$ is formally infinite, only a finite
number of terms in this sum contribute to a particular value of $m$,
so the sums in (\ref{D:14}) are finite.  Similar remarks apply to
$GH^{j-l-1}$.  The integral on the right side of (\ref{D:14}) is
\beq
I_{m'n'p}^{0,0,0} = 
\int s_{m'} (x) s_{n'} (x) {ds\over dx}  (x-p) dx .
\label{D:15}
\eeq
Note that the wavelet in (\ref{D:12}-\ref{D:13}) is replaced by a scaling 
function because equation (\ref{D:8}) expresses the scale $1/{2^l}$ 
wavelet as a linear combination of scale $1/2^{l+1}$ scaling functions.

Next we use translational invariance to transform the first index to zero 
\beq
I_{m'n'p}^{0,0,0} = I_{0,n'-m',p-m'}^{0,0,0}.
\label{D:16}
\eeq
Thus we can express any of the integrals,  $I_{mnp}^{k,k+l,k+j}$, in terms 
of the $(2K-2)^2$ non-trivial coefficients
$I_{0mn}^{0,0,0}$ .

Because of the compact support of the scaling function and its derivative,
$I_{0mn}^{0,0,0}$ is non-vanishing only for
values of $m,n$ satisfying 
\beq
-2K+1 < m,n < 2K-1;\quad  \vert m-n \vert < 2K-1 .
\label{D:18}
\eeq
These integrals have the form 
\beq
I_{0mn}^{0,0,0}  = \int s (x) s (x-m) {ds\over dx}  (x-n) dx .
\label{D:19}
\eeq
Next we use (\ref{D:7},\ref{D:8},\ref{D:9}) and (\ref{D:11})
and translational invariance to get the following scaling equation  
\[
I_{0mn}^{0,0,0} = {1 \over \sqrt{2}} I_{0mn}^{-1,-1,-1} = 
{2 \over \sqrt{2}}\sum_{l'm'n'}H_{ll'}H_{mm'}H_{nn'}I_{l'm'n'}^{000} =
\]
\beq 
{\sqrt{2}}\sum_{l'm'n'}H_{ll'}H_{mm'}H_{nn'}I_{0,m'-l',n'-l'}^{000}
\label{D:20}
\eeq
which is a system of $(2K-2) \times (2K-2)$ homogeneous equations 
for the values of $I_{0mn}^{0,0,0}$ with $-2K+1 < m,n < 2K-1$.   

In order to solve this system we also need an inhomogeneous equation, which 
must be related to the normalization condition on the scaling function. 
The Daubechies $K=3$ scaling function is defined so 
$1$, $x$ and $x^2$ can be pointwise expanded as linear combinations
of scaling functions 
\beq
1 = \sum _n a_n s_n (x)  \qquad 
\qquad
\mbox{and}
\qquad
x = \sum_n b_n s_n (x).
\label{D:22}
\eeq
The normalization condition (\ref{A:3}) and the orthogonality 
condition (\ref{A:12a}) give $a_n =1$ or 
\beq
1= \sum_n s_n (x), 
\label{D:23}
\eeq
which is a restatement of the partition of unity property.
To calculate $b_n$ use the unitarity of $D$ and the scaling 
equation to get  
\[
b_n = \int x s_n (x) dx = n + \int x s(x) dx =
n + \int D^{-1} x D^{-1} s (x) dx =  
\]
\[
n+ {1 \over 2\sqrt{2}} \sum_l h_l \int x s(x-l) dx =
\]
\[
n+ {1 \over 2\sqrt{2}} \sum_l l h_l \int s(x) dx +  
{1 \over 2\sqrt{2}} \sum_l h_l \int x s(x) dx =
\]
\beq
n+ {1 \over 2\sqrt{2}} \sum_l l h_l  +  
{1 \over 2} \int x s(x) dx , 
\label{D:24}
\eeq
where we used (\ref{A:9}) in (\ref{D:24}).
Setting $n=0$ gives 
\beq
\int x s(x) dx = {1 \over \sqrt{2}} \sum_l l h_l .
\eeq
It follows that 
\beq
b_n = n + {1 \over \sqrt{2}}\sum_l lh_l 
\label{D:26}
\eeq
and 
\beq
x = \sum_n n s_n (x) + {1 \over \sqrt{2}}\sum_l lh_l .
\label{D:26a}
\eeq
If we differentiate (\ref{D:26a}) we get 
a partition of unity for the derivative of the scaling function, 
\beq
1 = \sum n s_n'(x) . 
\label{D:27}
\eeq 
Equation (\ref{D:27}) implies that the 
coefficient functions satisfy
\beq
\sum_n n I_{0mn} = \delta_{m0}.
\label{D:29}
\eeq
This gives a non-trivial inhomogeneous equation when $m=0$.
If we sum over the first index we get the equivalent 
inhomogeneous equation
\beq
\sum_{mn} n I_{0mn} = 1.
\label{D:30}
\eeq
The next step is to identify a set of independent linear equations,
including (\ref{D:30}), for the non-zero $I_{0mn}^{0,0,0}$.  Solving
these equations gives the non-vanishing $I_{0mn}^{0,0,0}$.

Reversing these steps we get the following expression for the general 
coefficient in terms of the coefficients $I_{0mn}^{0,0,0}$:
\beq
I_{mnp}^{k,k+l,k+j}  = 
2^{-3(k+j)/2} \sum_{m'n'}  H^j_{mm'} (GH^{j-l-1})_{nn'}  
I_{0,n'-m',p-m'}^{0,0,0} .
\label{D:32}
\eeq
This shows that the scaling properties lead to explicit
formulas for an infinite number of coefficients in terms of the finite
set of coefficients $I_{0,m,n}^{0,0,0}$ with $-2K+1 < m,n,m-n < 2K-1$.

Given the above method to compute all of the integrals without powers
of $x$, it is easy to generalize the construction to treat integrands
involving powers of $x$. This computation uses the translations that
appear in the scaling equations to reduce the desired integral to one
of the types discussed above.

The basic trick is to combine (\ref{D:7}-\ref{D:11}). 
For example, all of the moments, 
\beq
<x^m>_{s} = \int s (x) x^m dx 
\qquad
<x^m>_{w} = \int w (x) x^m dx ,
\label{D:34}
\eeq
can be constructed recursively from the normalization
condition 
\beq
<x^0>_{s} = \int dx s (x) = 1
\label{D:35}
\eeq
using the scaling equation
\[
<x^m>_{s} = \int D^{-1}x^m  D^{-1}s (x) dx 
\]
\[
={1 \over \sqrt{2}}{1 \over 2^m} \sum_l h_l \int x^m s(x-l) dx   
\]
\[
={1 \over \sqrt{2}}{1 \over 2^m} \sum_l h_l \int (x+l)^m  s(x)dx 
\]
\beq
={1 \over \sqrt{2}}{1 \over 2^m} \sum_l h_l \sum_{k=0}^m
{ m! \over k! (m-k)!} l^{m-k} 
\int x^k s(x) dx .
\label{D:36}
\eeq
Using $\sum_{l} h_l = \sqrt{2}$, and moving the $k=m$ term to the left side 
of equation (\ref{D:36}) gives the recursion relation for the $m$-th
moments in terms of $k<m$ moments
\beq
\int x^m s(x) dx =  {1 \over 2^m-1}{ 1 \over \sqrt{2}}  \sum_{k=0}^{m-1}
{ m! \over k! (m-k)!} \left (\sum_{l=1}^{2K-1} h_l l^{m-k}\right ) 
\int x^k s(x) dx . 
\label{D:37}
\eeq
Moments of the wavelets are obtained by replacing $h_l$ in (\ref{D:36}) 
by $g_l$:
\[
<x^m>_{w} = \int w (x) x^m dx =
\]
\beq 
={1 \over \sqrt{2}}{1 \over 2^m} \sum_l g_l \sum_{k=0}^m
{ m! \over k! (m-k)!} l^{m-k} 
\int x^k s(x) dx .
\label{D:38}
\eeq
This means the first moment can be expressed in terms of the $0^{th}$
moment, which is $1$ by the scale fixing condition.  This method can
be generalized to calculate moments multiplied by arbitrary products
of scaling functions, wavelets and their first derivatives.  In the
general case one gets a linear system for the k$^{th}$ moment in terms of the
$0,1,\cdots, k-1^{st}$ moments.

Using these methods all of the coefficients appearing in the expressions
for the Poincar\'e generators can be expressed in terms of the scaling 
coefficients, $h_l$, and the solution of a finite system of linear equations.
The factors of $H^j$ and $GH^{j-l-1}$ in (\ref{D:32}) are dominated by the 
terms involving $j$ or $j-l-1$ products of $h_1=g+4\approx 1.1$, or
$H_{mn} \to 1.1 \delta_{1-n-2m,0}$ and   $G_{mn} \to 1.1 \delta_{4-n-2m,0}$.

Calculations of (\ref{C:35}-\ref{C:37}) are done explicitly in the appendix.

These methods can be used to calculate all of the factors that appear
in the formal expressions for the Poincar\'e generators.  The scaling 
properties of these coefficients can be read off using (\ref{D:2}-\ref{D:5}).  
For the coefficients involving only the scale $k$ scaling functions 
these coefficients have the following scaling properties:
\beq
D^k_{\mathbf{m}\mathbf{n}}={2^{2k}} 
D^0_{\mathbf{m}\mathbf{n}} 
\label{D:39}
\eeq
\beq
\mathbf{P}^{k}_{\mathbf{m},\mathbf{n}}= 2^k 
\mathbf{P}^{0}_{\mathbf{m},\mathbf{n}} 
\label{D:40}
\eeq
\beq
\mathbf{J}^{k}_{\mathbf{m},\mathbf{n}}=
\mathbf{J}^{0}_{\mathbf{m},\mathbf{n}}
\label{D:41}
\eeq
\beq
\mathbf{F}^k_{\mathbf{n}\mathbf{m}} = {1 \over 2^k}
\mathbf{F}^0_{\mathbf{n}\mathbf{m}}  
\label{D:42}
\eeq
\beq
\mathbf{G}^k_{\mathbf{n}\mathbf{m}} ={2^{k}}
\mathbf{G}^0_{\mathbf{n}\mathbf{m}}.
\label{D:43}
\eeq

\section{Poincar\'e Symmetry}

There are three properties of the wavelet basis that are useful for
investigating Poincar\'e symmetry breaking in discrete truncations of
field theory.  First, the scaling functions on a fixed scale are a
compactly supported, locally finite partition of unity.  This makes it
possible to test the Poincar\'e symmetry in finite volumes.  Second,
the scaling functions and wavelets form a basis, which means that it
is possible to precisely identify the corrections that are needed to
restore Poincar\'e invariance in truncated theories.  Third, because
it is possible to compute the coefficients of the operators that
restore the commutation relations to a set of truncated generators, it
is possible to identify and classify the size of all of the correction
terms.

In this section we discuss these three properties in more detail.  First
we consider the partition of unity property.  Up to an overall
multiplicative constant the scaling functions on a given scale, $1/2^k$,
form a partition of unity that has the form
\beq
1 = \sum_{\mathbf{m}} 2^{-3k/2} s^k_{\mathbf{m}}(x) .
\label{E:4}
\eeq
While this sum is formally infinite, at any given point there are at
most $(2K-2)^3$ non-vanishing terms in this sum.

The partition of unity can be utilized to decompose each of the
Poincar\'e generators into sums of localized operators.  To do this
note that Noether's theorem gives formal expressions for the
Poincar\'e generators as integrals of local densities
(\ref{B:9}-\ref{B:10}) over the $t=0$ surface.  The local densities
involve products of field operators and their derivatives at
the same spacetime point.

The local density, $O^i(x)$, of the $i$-th Poincar\'e generator
satisfies  
\beq
[ O^i(\mathbf{x},0) , O^j(\mathbf{y},0) ] = i \delta (\mathbf{x}-\mathbf{y}) 
f^{ijk}O^k (\mathbf{y},0) 
\label{E:1}
\eeq
where the $f^{ijk}$ are the structure constants of the Poincar\'e group. 
Generators are integrals of these densities over the $t=0$ hyperplane 
\beq
O_i := \int d\mathbf{x} O^i(\mathbf{x},0).  
\label{E:2}
\eeq
These integrals are the conserved charges in Noether's theorem and
formally become time independent after integrating over a fixed-time
hyperplane.  The Poincar\'e commutation relations 
are obtained by integrating both sides of equation (\ref{E:1})
over a fixed-time hyperplane.  This gives the commutation relations
\beq
[ O^i , O^j ] = i  f^{ijk}O^k \qquad O^i = \int d\mathbf{x} O^i(\mathbf{x},0) .
\label{E:3}
\eeq

While the generators are not local operators, they involve integrals of 
local densities over a 3-dimensional hyperplane, and
the partition of unity for scale-$1/2^k$ scaling functions can be used 
to express the commutation relations in an almost local form.
To do this we insert the partition of unity, (\ref{E:4}), in 
(\ref{E:2}) to get  
\beq
O^i = \sum_{\mathbf{m}} \int O_{\mathbf{m}}^{ki} 
\label{E:5}
\eeq
where the smeared density is 
\beq
O_{\mathbf{m}}^{ki} := \int 2^{-3k/2} s^k_{\mathbf{m}}(\mathbf{x}) 
O^i (\mathbf{x},0) d \mathbf{x}
\label{E:6}
\eeq
and the integrals are over the $t=0$ hyperplane. 

In this notation the exact commutation relations (\ref{E:3})
can be expressed in terms of the sums:
\beq
\sum_{\mathbf{m}\mathbf{n}} [ O^{ki}_{\mathbf{m}} , O^{kj}_{\mathbf{n}} ] 
= i \sum_l f^{ijl}\sum_{\mathbf{n}} O^{kl}_{\mathbf{n}}.  
\label{E:7}
\eeq
While these sums have an infinite number of terms, there are no
convergence problems because when (\ref{E:7}) operates on the dense
set of states $\vert \Psi \rangle$ constructed out of finite linear
combinations of products of creation operators smeared against
wavelets and scaling functions, whose support is necessarily contained
in a bounded region, only a finite number of terms in the sum
(\ref{E:7}) are non-zero, giving
\beq
\sum_{\mathbf{m}\mathbf{n}} [ O^{ki}_{\mathbf{m}} , O^{kj}_{\mathbf{n}} ]
\vert \psi \rangle 
= i \sum_l f^{ijl}\sum_{\mathbf{n}} O^{kl}_{\mathbf{n}} \vert \psi \rangle . 
\label{E:9}
\eeq
In (\ref{E:9}) the sums are finite, with $\mathbf{m}$ and $\mathbf{n}$
running only over the indices corresponding to scaling functions in
the partition of unity that have common support with the smearing
functions $s_{\mathbf{n}}^k( \mathbf{x})$ and $w_{\mathbf{j}}^l(
\mathbf{x})$ used in the discrete creation operators.

Equation (\ref{E:9}) is an exact consequence of Poincar\'e invariance, 
and if it is satisfied on the dense set of vectors generated by finite 
numbers of discrete creation operators, one recovers exact Poincar\'e 
invariance.

Thus Poincar\'e invariance can be tested by replacing the infinite
sums in (\ref{E:7}) by a suitable finite sum of the localized operators,
$O^{kl}_{\mathbf{n}}$.  Where approximations enter is that the local
densities $O^i (\mathbf{x},0)$ that appear in the definition of
$O^{kl}_{\mathbf{n}}$ are constructed out of normal products of field
operators.  The fields that appear in these expressions can all be
expanded in the form (\ref{C:1}-\ref{C:2}). The partition of unity projects
the product of different scale discrete fields on the scale $1/2^k$
subspace.  As long as there is no truncation everything is exact.  The
problem is that for normal products of more than one field, there are
scale $1/2^{k}$ contributions to the operator products that involve
fields on scales smaller than $1/2^k$.  If we truncate the theory by
discarding all of contributions to the field operators associated with
scales smaller than $1/2^k$, then the discarded scale $1/2^k$
contributions that come from projection of products of smaller scale
field operators are precisely the corrections that are needed to
restore Poincar\'e invariance.

To understand the structure of these discarded terms consider the
exact expression for the linear momentum of a free field using the
scale $1/2^k$ scaling functions and associated wavelets.  The scalar
density $\mathbf{O}(\mathbf{x},0)$ for the momentum operator is (\ref{B:12}) 
which is the 
normal product the $\Pi(x)$ field and the gradient of the
$\Phi(x)$ field:  
\beq
\mathbf{P}(\mathbf{x}) = 
- : \Pi (\mathbf{x},0)\pmb{\nabla} \Phi (\mathbf{x},0):.
\label{E:10}
\eeq
Using the partition of unity (\ref{E:4}) equation 
(\ref{B:12}) can be expressed 
exactly as
\beq
\mathbf{P}= \sum_{\mathbf{m}} \mathbf{P}^k_{\mathbf{m}}
\label{E:11}
\eeq
where 
\beq
\mathbf{P}^k_{\mathbf{m}} = 
- \int : \Pi (\mathbf{x},0)\pmb{\nabla} \Phi (\mathbf{x},0): 
2^{-3k/2} s^k_{\mathbf{m}}(x) d \mathbf{x} .
\label{E:12}
\eeq

Alternatively, we can express $\mathbf{P}^k_{\mathbf{m}}$ using the 
exact expansion of each of the fields in the scale $1/2^k$ wavelet basis 
and smearing the result with the scale $1/2^k$ partition of unity: 
\[
\mathbf{P}^{k}_{\mathbf{m}}=
\sum_{\mathbf{j},\mathbf{n}} :\Pi^{k}(\mathbf{j},0) P^{kkk}
_{\mathbf{m}\mathbf{n}\mathbf{j}}
\Phi^k(\mathbf{n},0): + 
\]
\[
\sum_{l,\alpha, \mathbf{j},\mathbf{n}}:\Pi^{l}(\mathbf{j},\alpha,0) 
P^{lkk}_{\mathbf{j},\alpha,\mathbf{n}\mathbf{m}}
\Phi^k(\mathbf{n},0): +
\]
\[
\sum_{l,\alpha, \mathbf{j},\mathbf{n}}:\Pi^{k}(\mathbf{n},0) 
P^{klk}_{\mathbf{j},\mathbf{n},\alpha,\mathbf{m}}
\Phi^l(\mathbf{n},\alpha,0): + 
\]
\beq
\sum_{l,\alpha, \mathbf{j},\mathbf{n}}:\Pi^{l}(\mathbf{n},\alpha,0) 
P^{lsk}_{\mathbf{j},\alpha,\mathbf{n},\beta,\mathbf{m}}
\Phi^s(\mathbf{n},\beta,0): 
\label{E:13}
\eeq
where the expansion coefficients are 
\beq
P^{kkk}
_{\mathbf{m}\mathbf{n}\mathbf{j}} = 
- \int s^k_{\mathbf{m}}(\mathbf{x})  s^k_{\mathbf{n}} (\mathbf{x})
 \pmb{\nabla}s^k_{\mathbf{j}}(\mathbf{x}) d\mathbf{x} 
\label{E:14}
\eeq
\beq
P^{lkk}_{\mathbf{j},\alpha,\mathbf{n}\mathbf{m}}=
- \int w^l_{\mathbf{j},\alpha}(\mathbf{x}) s^k_{\mathbf{n}}(\mathbf{x})   
\pmb{\nabla}s^k_{\mathbf{m}}(\mathbf{x}) d\mathbf{x} 
\label{E:15}
\eeq
\beq
P^{klk}_{\mathbf{j},\mathbf{n},\alpha,\mathbf{m}}=
- \int s^k_{\mathbf{m}}(\mathbf{x})s^k_{\mathbf{j}} (\mathbf{x})
\pmb{\nabla}w^l_{\mathbf{n},\alpha}(\mathbf{x}) d\mathbf{x} 
\label{E:106}
\eeq
\beq
P^{lsk}_{\mathbf{j},\alpha,\mathbf{n},\beta,\mathbf{m}}=
- \int w^l_{\mathbf{j},\alpha}(\mathbf{x})
s^k_{\mathbf{m}}(\mathbf{x})  \pmb{\nabla}w^s_{\mathbf{n},\beta}(\mathbf{x})d\mathbf{x}. 
\label{E:17}
\eeq
The coefficients above are numerical coefficients.  

To understand the interpretation of these results we note that 
(\ref{E:9}) is an exact consequence of the commutation relations
expressed in terms of the scale $1/2^k$ partition of unity. 
The expansion (\ref{E:13}) is exact, and the resulting operators
satisfy (\ref{E:9}).  If the fields are replaced by the resolution 
$1/2^k$ approximations given by the first term on the right-hand side of 
(\ref{C:1}) and (\ref{C:2}), then (\ref{E:13}) would be replaced by 
an ``approximation'' consisting of only the first line of (\ref{E:13}). 
This approximation would violate the commutation relations (\ref{E:9}). 
 
The ``corrections'' defined by the second through fourth line of
(\ref{E:13}) provide the missing physics from smaller scales that is
needed to restore the commutation relations.  Note that (\ref{E:9})
only involves operators that act in a finite volume, so any problems
due to a volume truncation are not relevant.

In the free-field case the generators (\ref{B:19}-\ref{B:22}) are sums
of normal products of one creation and one annihilation operator.  The
commutator of two operators with the structure
$a^k(\mathbf{m})^{\dagger} a^k(\mathbf{n})$ leads to another operator
of the same form.  In addition commutators of operators of the
structure $a^k(\mathbf{m})^{\dagger} b^l(\mathbf{n},\alpha)$ with
$b^l(\mathbf{n},\alpha)^{\dagger} a^k(\mathbf{j})$ also lead to
structures of the form $a^k(\mathbf{m})^{\dagger} a^k(\mathbf{j})$.
 
The breaking of Poincar\'e invariance is related to the 
question of how large are the corrections
$P^{lkk}_{\mathbf{j},\alpha,\mathbf{n}\mathbf{m}} $,
$P^{klk}_{\mathbf{j},\mathbf{n},\alpha,\mathbf{m}}$ and
$P^{lsk}_{\mathbf{j}\alpha \mathbf{n}\beta \mathbf{m}}$ that restore
the commutation relations relative to the coefficients
$\mathbf{P}^{kkk}_{\mathbf{m}\mathbf{n}\mathbf{j}}$.  that define the
approximate generators in a model with a resolution $1/2^k$ cutoff.

These coefficients can all be computed exactly using the methods of
the previous section in order to identify the largest correction
terms.  

\section{Renormalization Group}

The wavelet basis decomposes the Hilbert space $L^2(\mathbb{R}^3)$
into an infinite orthogonal direct sum of infinite-dimensional subspaces of
successively finer resolution.

The Hamiltonian can be truncated at any resolution.  The resulting operator
is similar to a Hamiltonian with integrals over fields replaced by 
sums of fields averaged over lattice blocks, except the averaging
functions are products of scaling functions of a given scale.
In our free-field example the resolution $1/2^k$ truncated Hamiltonian 
has the form $H(a)$ in (\ref{C:32}).

The scaling properties of the wavelet basis make it natural for
implementing renormalization group transformations.  For the wavelet
realization one starts with a Hamiltonian truncated to a given fixed
scale, $1/2^k$.  The truncated Hamiltonian is obtained from the formal
expansion for the ``exact'' Hamiltonian by discarding contributions
from fields associated with scales smaller than $1/2^k$.  In practice
there will also be a volume cutoff.

The truncated Hamiltonian defines a class of resolution $1/2^k$
Hamiltonians depending on the choice of bare parameters $\lambda_{01}
\cdots \lambda_{0n}$.  The bare parameters can be initially fixed by
computing $n$ scale $1/2^k$ observables, $O^k_i(\lambda_{01}, \cdots ,
\lambda_{0n})$, and adjusting the bare parameters so they agree with
fixed ``experimental'' values, $Q^{k*}_i$, of these observables 
\beq
O^k_i(\lambda^*_{01} \cdots \lambda^*_{0n}) = O^{k*}_i 
\qquad 1\leq i \leq n .
\label{F:0}
\eeq
We assume that these equations can be solved and we denote the solution
of (\ref{F:0}) by $\lambda^*_{10}(k), \cdots , \lambda^*_{n0}(k)$,
where the factor $k$ indicates the resolution of the truncated
Hamiltonian.

The next step of the renormalization program is to consider the class
of truncated Hamiltonians at resolution $1/2^{k+1}$; i.e. increase the
resolution by a factor of 2.  These Hamiltonians are still functions
of the same $n$ bare parameters.  Using (\ref{A:21}) in (\ref{C:3})
and (\ref{C:5}) the fields can be expressed in terms of the scale
$1/2^k$ scaling function fields and the scale $1/2^k$ wavelet fields.
Eliminating the wavelet field degrees of freedom, results in a set of
effective Hamiltonians in the $1/2^k$ resolution degrees of freedom
that include the effects of the eliminated resolution $1/2^{k+1}$ degrees
of freedom.  The resulting effective Hamiltonians are more complicated, 
but they are still parameterized by the same bare parameters. 
One can again solve (\ref{F:0}) with the same observables associated 
with the new effective Hamiltonian.  This leads to 
new values of the bare parameters by $\lambda^*_{10}(k+1), \cdots,
\lambda^*_{n0}(k+1)$.

We can repeat this process starting with the effective Hamiltonians
with scale $1/2^k$ degrees of freedom that include the effects of the
eliminated scale $1/2^{k+1}$ degrees of freedom to construct a new
effective Hamiltonians with scale $1/2^k$ degrees of freedom that
include the effects of the eliminated scale $1/2^{k+2}$ degrees of
freedom.  Solving (\ref{F:0}) again leads to a new set of bare 
parameters $\lambda^*_{0}(k+2), \cdots \lambda^*_{n0}(k+2)$.  This 
process can be continued, incorporating physics from successively 
smaller scales in scale $1/2^k$ Hamiltonian.

In this process it useful to work with a volume cutoff of fixed size
that is large enough to not impact the value of the scale $1/2^k$
``experimental'' observables that determine the sequence of bare
coupling constants.

In general the bare parameters $\lambda^*_{0}(l)$ will diverge or
vanish in the limit that $1/2^{k+l} \to 0$.  Renormalization expresses
the bare parameters $\lambda^*_{m0}(k+l)$ in terms of the renormalized
parameters $\lambda^*_{m}(k+l)$ multiplied by $1/2^{k+l}$-dependent
factors that remove the leading scale dependence from the bare
parameters.  For power-law behavior, the renormalized and
unrenormalized parameters are related by $\lambda^*_{m}(k+l) =
(2^{l})^{s_m} \lambda^*_{m0}(k+l)$, where the exponents $s_m$ are
chosen to remove the leading scale dependence from 
$\lambda^*_{m0}({k+l})$ as $l\to \infty$.

If the renormalized parameters converge to finite fixed points in the
high-resolution limit, then the resulting scale $1/2^k$ effective
Hamiltonian is equivalent to one obtained from the renormalized
Hamiltonian by exactly eliminating high-resolution degrees of freedom,
where both the full and effective Hamiltonians have finite
renormalized parameters.

In what follows we discuss the role of the wavelet basis 
in the formulation of the renormalization group transformations
that determine the dependence of the bare coupling constants on 
resolution of the eliminated degrees of freedom.  

We begin by considering the relation of the free Hamiltonians, $H(a)$,
truncated on two different scales.  To do this we use the scaling
equation in the form (\ref{A:16}) to obtain the following identities:
\[
\delta_{mn} = \int s_{m}^k (x) s_n^k(x) dx = 
\sum_{jl} h_{l-2m}h_{j-2n}\int s_l^{k+1}(x)s_j^{k+1}(x)dx =
\]
\beq  
\sum_{jl} h_{l-2m}h_{j-2n}\delta_{jl} =
\sum_{j} h_{j-2m}h_{j-2n}  
\label{F:1}
\eeq
and 
\[
D^k_{mn}= \int s^{k\prime}_m (x)s^{k\prime}_n (x) dx = 
\]
\beq
\sum_{jl} h_{l-2m}h_{j-2n}\int s_l^{k+1\prime}(x)s_j^{k+1\prime }(x)dx = 
\sum_{jl} h_{l-2m}h_{j-2n}D^{k+1}_{lj} .
\label{F:2}
\eeq

Next we consider the individual terms in the resolution $1/2^{k+1}$
truncated free field Hamiltonain $H^{k+1}(a)$.  This Hamiltonain 
is the sum of the following three terms 
\beq
{1 \over 2} \sum_{\mathbf{n}} :{\Pi}^{k+1}(\mathbf{n},0){\Pi}^{k+1}
(\mathbf{n},0):,
\label{F:3}
\eeq
\beq
\sum_{\mathbf{m}\mathbf{n}}:\Phi^{k+1} (\mathbf{m} ,0) D^{k+1}_{\mathbf{m}\mathbf{n}}
\Phi^{k+1} (\mathbf{n},0):,
\label{F:4}
\eeq
and 
\beq
\mu^2 \sum_{\mathbf{n}} :{\Phi}^{k+1}(\mathbf{n},0){\Phi}^{k+1}(\mathbf{n},0):).
\label{F:5}
\eeq
The discrete fields in these terms have the form 
\beq
{\Phi}^{k+1}(\mathbf{n},0) = 
\int \Phi (\mathbf{x},0) s^{k+1}_{n_1}(x)
s^{k+1}_{n_2}(y) s^{k+1}_{n_3}(z)d\mathbf{x} 
\label{F:6}
\eeq
and
\beq
{\Pi}^{k+1}(\mathbf{n},0) = 
\int \Pi (\mathbf{x},0) s^{k+1}_{n_1}(x)s^{k+1}_{n_2}(y) 
s^{k+1}_{n_3}(z)d\mathbf{x}. 
\label{F:7}
\eeq
Using (\ref{A:18}) these can be expressed in the form  
\[
{\Pi}^{k+1}(\mathbf{n},0) = 
\]
\[
\int \Pi (\mathbf{x},0) s^k_{n_1}(x_1)s^k_{n_2}(x_2) s^k_{n_3}(x_3)d\mathbf{x}  =
\]
\[
\int \Pi (\mathbf{x},0) 
(\sum_{j_1} h_{n_1-2j_1} s^{k}_{j_1}(x_1) +  \sum_{j_1} g_{n_1-2j_1} w^{k-1}_{j_1}(x_1))
\times
\]
\beq
(\sum_{j_2} h_{n_2-2j_2} s^{k}_{j_2}(x_2) +  \sum_{j_2} g_{n_2-2j_2} w^{k-1}_{j_2}(x_2))
(\sum_{j_3} h_{n_3-2j_3} s^{k}_{j_3}(x_3) +  \sum_{j_3} g_{n_3-2j_3} w^{k-1}_{j_3}(x_3))
d\mathbf{x} 
\label{F:8}
\eeq
\[
{\Phi}^{k+1}(\mathbf{n},0) = 
\]
\[
\int \Phi (\mathbf{x},0) s^k_{n_1}(x_1)s^k_{n_2}(x_2) s^k_{n_3}(x_3)d\mathbf{x}  =
\]
\[
\int \Phi (\mathbf{x},0) 
(\sum_{j_1} h_{n_1-2j_1} s^{k}_{j_1}(x_1) +  \sum_{j_1} g_{n_1-2j_1} w^{k-1}_{j_1}(x_1))
\times
\]
\beq
(\sum_{j_2} h_{n_2-2j_2} s^{k}_{j_2}(x_2) +  \sum_{j_2} g_{n_2-2j_2} w^{k-1}_{j_2}(x_2))
(\sum_{j_3} h_{n_3-2j_3} s^{k}_{j_3}(x_3) +  \sum_{j_3} g_{n_3-2j_3} w^{k-1}_{j_3}(x_3))
d\mathbf{x} .
\label{F:9}
\eeq
These expressions have the form 
\beq
{\Phi}^{k+1}(\mathbf{n},0) = \sum_{\mathbf{j}}
\prod_i h_{n_i-2j_i}{\Phi}^k(\mathbf{j},0) + \cdots
\label{F:10}
\eeq
\beq
{\Pi}^{k+1}(\mathbf{n},0) = \sum_{\mathbf{j}}
\prod_i h_{n_i-2j_i}{\Pi}^k(\mathbf{j},0) + \cdots
\label{F:11}
\eeq
where the $\cdots$ terms represent terms where at least one scale 
$k$ wavelet $w^k_n(x)$ appears in the integral.  Using these
expressions in (\ref{F:3}) gives
\[
{1 \over 2} \sum_{\mathbf{n}} :{\Pi}^{k+1}(\mathbf{n},0){\Pi}^{k+1}
(\mathbf{n},0): = 
\]
\[
{1 \over 2} 
(\sum_{\mathbf{n}} :\prod_i h_{n_i-2j_i}{\Pi}^k(\mathbf{j},0)
\prod_m h_{n_m-2r_m}{\Pi}^k(\mathbf{r},0): +\cdots )=
\]
\beq
{1 \over 2} 
(\sum_{\mathbf{n}} :{\Pi}^k(\mathbf{j},0){\Pi}^k(\mathbf{j},0) +\cdots ) =
\label{F:12}
\eeq
where we have used (\ref{F:1}) three times.  We see that this term in
the resolution $1/2^{k+1}$ Hamiltonian can be expressed as the
corresponding term in the resolution $1/2^k$ Hamiltonian plus wavelet
related corrections that restore the full $1/2^{k+1}$ scale physics.

Similarly we find the mass term has the same form
\beq
\mu^2 \sum_{\mathbf{n}} :{\Phi}^{k+1}(\mathbf{n},0){\Phi}^{k+1}(\mathbf{n},0):)=
\mu^2 \sum_{\mathbf{n}} :{\Phi}^{k}(\mathbf{n},0){\Phi}^{k}(\mathbf{n},0): + \cdots .
\label{F:13}
\eeq
Finally we consider the term (\ref{F:4}) 
\[
\sum_{\mathbf{m}\mathbf{n}}:\Phi^{k+1} (\mathbf{m} ,0) D^{k+1}_{\mathbf{m}\mathbf{n}}
\Phi^{k+1} (\mathbf{n},0): 
=
\]
\[
\sum_{\mathbf{m}\mathbf{n}}:
(\sum_{\mathbf{j}}
\prod_i h_{m_i-2j_i}{\Phi}^k(\mathbf{j},0) + \cdots) 
\times 
\]
\beq
\sum  D^{k+1}_{\mathbf{m_i}\mathbf{n_i}}
(\sum_{\mathbf{l}}
\prod_i h_{n_i-2l_i}{\Phi}^k(\mathbf{l},0) + \cdots)
\label{F:14}
\eeq
Using (\ref{F:1}) and (\ref{F:2}) this becomes 
\beq
\sum_{\mathbf{m}\mathbf{n}}:\Phi^{k} (\mathbf{m} ,0) D^{k}_{\mathbf{m}\mathbf{n}}
\Phi^{k} (\mathbf{n},0): + \cdots .
\label{F:15}
\eeq

{\it Taken together these calculations show that the resolution
  $1/2^{k+1}$ free Hamiltonain is equal to the resolution $1/2^{k}$
  free Hamiltonian plus a correction that fills in the missing degrees
  of freedom that appear in the scale $1/2^{k+1}$ but not on scale
  $1/2^k$.}  The correction terms all involve fields smeared with
wavelets.  An important observation is that for free fields the
derivative terms provide the coupling between the two scales.  A
similar analysis can be used to show that all of the free-field
Poincar\'e generators have the property that the resolution
$1/2^{k+1}$ operators are equal to the resolution $1/2^k$ generators
plus wavelet field corrections.

While we derived these results by considering the example of a free field,
the result also holds for interactions.  For example a scale $1/2^{k+1}$
truncated $\int :\Phi^n(\mathbf{x},0):d \mathbf{x}$ 
interaction has the form 
\beq
\sum_{m_1 \cdots m_n} :\Phi^{k+1}(\mathbf{m_1},0)
\cdots \Phi^{k+1}(\mathbf{m_n},0): \Gamma^{k+1}_{\mathbf{m}_1 \cdots \mathbf{m}_n}
\label{F:16}
\eeq
where
\beq
\Gamma^{k+1}_{\mathbf{m}_1 \cdots \mathbf{m}_n} =
\int s^{k+1}_{\mathbf{m}_1}  (\mathbf{x}) \cdots s^{k+1}_{\mathbf{m}_n}  
(\mathbf{x}) d\mathbf{x} .
\eeq
It is straightforward to show, 
using (\ref{A:16}) and (\ref{A:18}),
that
\beq
\Gamma^{k}_{\mathbf{m}_1 \cdots \mathbf{m}_n} =
\prod h_{l_{1i}-2m_{1i}} \cdots h_{l_{ni}-2m_{ni}} 
\Gamma^{k+1}_{\mathbf{l}_1 \cdots \mathbf{l}_n}
\eeq
which when used with (\ref{F:10}) and (\ref{F:11}) gives 
\[
\sum_{m_1 \cdots m_n} :\Phi^{k+1}(\mathbf{m_1},0)
\cdots \Phi^{k+1}(\mathbf{m_n},0) \Gamma^{k+1}_{\mathbf{m}_1 \cdots \mathbf{m}_n}=
\]
\beq
\sum_{m_1 \cdots m_n} :\Phi^{k}(\mathbf{m_1},0)
\cdots \Phi^{k}(\mathbf{m_n},0) \Gamma^{k}_{\mathbf{m}_1 \cdots \mathbf{m}_n}
+ \cdots
\label{F:17} 
\eeq
where the $\cdots$ terms represent the contributions where at least one
of the expansions functions is a scale $1/2^k$ wavelet, $w^k_m(x)$.

It follows that a resolution 
$1/2^{k+1}$ Hamiltonian has the form 
\beq
H^{k+1}(a) = H^k(a) + H^k(b)  + H^k (ab)    
\label{F:18}
\eeq
where $H^k(a)$ is the part of $H^{k+1}$ that has only scale $1/2^k$
scaling function fields, $H^k(b)$ has only scale $1/2^k$ wavelet fields
and $H^k (ab)$ contains the terms with at least one wavelet and one scaling 
function field.  In the interacting case the interaction also contributes
to $H^k(ab)$

The creation operators associated with scale $1/2^k$ scaling function
fields and the creation operators associated with at least one scale
$1/2^k$ wavelet field each generate mutually orthogonal subspaces on
the Fock space generated by the scale $1/2^{k+1}$ scaling function 
creation operators.

One can then seek a unitary transformation that Block diagonalizes
$H^{k+1}(a)$ on the subspace generated by the scale $1/2^k$ 
scaling-function creation operators.  The resulting effective Hamiltonain will
be a Hamiltonian in the scale $1/2^k$ degrees of freedom that includes
the effects of the eliminated scale $1/2^{k+1}$ degrees of freedom.
Both Hamiltonians are different functions of the same coupling constants
and mass parameters.  Renormalization of the parameters in the Hamiltonian 
of the model is necessary to keep the resolution $1/2^k$ observables fixed.  
This gives a new coarse-scale Hamiltonain that includes additional degrees
of freedom at a finer scale.

Because the resolution $1/2^{k}$ Hamiltonain has the same form for any
$k$, the starting scale is arbitrary so we can repeat this process,
successively eliminating degrees of freedom associated with smaller
and smaller scales, renormalizing the parameters in the Hamiltonain at
each step.  At some point one can stop and the result will be an
effective theory that describes the scale $1/2^{k}$ degrees of
freedom, including the effects of the eliminated scale $1/2^{k+l}$ degrees
of freedom, or one can proceed to try to find a fixed point of this
renormalization group equation.
 
The are a number of possible approaches that can be used to eliminate
the scale $1/2^{k+l}$ degrees that appear in the scale $1/2^{k+1}$
Hamiltonain but not in the scale $1/2^k$ Hamiltonian.

One method is to use the similarity renormalization group
method\cite{perry}.  This involves solving the differential equation
\beq
{dH(\lambda) \over d \lambda} = 
[H(\lambda)[H(\lambda, H^k(a)+H^k(b)] ] 
\label{F:19}
\eeq
with initial condition 
\beq
H(0) = H^{k+l}(a) = H^{k}(a)+H^k(b) + H^k(ab).   
\label{F:20}
\eeq
The resulting $H(\lambda)$ will evolve to a Hamiltonian that 
does not have coupling terms of the form $H(a,b)$.  In
principle this can also be applied to the full Hamiltonian.  It is easy
to see the iterative solution of this equation generates more
complicated interactions in the Hamiltonian with each iteration.  
The iteration involves commutators of discrete creation and annihilation 
operators multiplied by algebraically computable coefficients.

A second approach is to try to do this perturbatively.  In this
case the Hamiltonian is expressed in the form 
\beq
\left (
\begin{array}{cc}
H(a) & H^I \\
(H^I)^\dagger & H^c(a) 
\end{array}
\right ) 
\label{F:21}
\eeq
where in this expression $H(a)$ is the part of Hamiltonian that maps the 
Fock space generated by the scale $1/2^k$ scaling function creation operators
into itself,
$H^c$ is the projection of the resolution $1/2^{k+1}$  Hamiltonian on 
the orthogonal complement of this space, and $H^I$ and $(H^I)^{\dagger}$ 
are the parts of the resolution $1/2^{k+1}$  Hamiltonian that couple 
these two spaces.  

A unitary transformation of the Okobu  
\cite{okobu}\cite{glockle} form
\beq
U = \left (
\begin{array}{cc}
(I_a + A^{\dagger} A )^{-1/2} & -A^{\dagger} (I_b + A A^{\dagger} )^{-1/2} \\
A(I_a + A^{\dagger} A )^{-1/2} &  (I_b + A A^{\dagger} )^{-1/2}
\end{array}
\right ) 
\label{F:22}
\eeq
with 
\beq
A = \left (
\begin{array}{cc}
0 & 0 \\
A(ba) & 0 
\end{array}
\right ) 
\label{F:23}
\eeq
will block diagonailze the Hamiltonian on the Fock space generated by the
scale $1/2^k$ creation operators provided $A$ 
satisfies
\beq
A H(a) - H^c(a) A + (H^I)^{\dagger} - A H^I A =0.
\label{F:24}
\eeq 
Equation (\ref{F:24}) is a non-linear equation for $A$
that can be solved perturbatively in $H^I$.
In this case we seek a solution for $\lambda =1$ of 
\beq
A H(a) - H^c(a) A + \lambda (H^I)^{\dagger} - \lambda A H^I A =0.
\label{F:25}
\eeq
with 
\beq
A = \sum_{n=1}^\infty \lambda^n A_n
\label{F:26}
\eeq
An important feature of the wavelet method is that the coupling of the 
scales proceeds through the terms $H(ab)$.  These terms involve 
discrete operators multiplied by coefficients that involve integrals 
over product of scaling functions, wavelets and their derivatives.
Since these coefficients can all be computed exactly (using finite linear
algebra) it is possible to identify the largest or most important 
terms and discard the smaller ones to get more efficient approximate
solutions.

   
\section{Gauge Invariance}

To motivate the implementation of local gauge invariance in the
wavelet representation of field theory we consider the example of an
$SU(3)$ gauge field.  The treatment of full gauge invariance with
respect to fields smeared over four space-time variables or
time-independent gauge transformation restricted to a fixed-time
hyperplane for fields smeared over a basis for the hyperplane is
similar.  Since most of our development has been for fields smeared
over a basis for the fixed-time hyperplane, we limit our discussion of
gauge transformations to time-independent gauge transformations.

In the wavelet basis the basis functions are not local, however if all
of the basis functions are retained linear combinations can be used to
describe observables associated with arbitrarily small
regions. Locally independent gauge transformations can be built out of
independent gauge transformation associated with each of the
individual discrete wavelet fields.  To be specific we consider an
$SU(3)$ color gauge group.

We consider transformations of the form
\beq
\Phi^k_c(\mathbf{n},t) \to \Phi^{\prime k}_c(\mathbf{n},t)
=
\sum_{c'} V^k_{cc'}(\mathbf{n},t)  \Phi^k_{c'}(\mathbf{n},t)
\label{G:1}
\eeq
\beq
\Phi^l_c(\mathbf{n},\alpha, t) \to \Phi^{\prime l}_c(\mathbf{n},\alpha, t)
= \sum_{c'} V^l_{cc'}(\mathbf{n},\alpha,t)  \Phi^l_{c'}(\mathbf{n},\alpha, t)
\label{G:2}
\eeq
where 
$V^k_{cc'}(\mathbf{n},t)$ and $V^l_{cc'}(\mathbf{n},\alpha,t)$ are 
$SU(3)$ valued functions of the various parameters.  They are independent 
for each independent discrete field operator.

It is obvious that quadratic expressions of the form: 
\beq
\sum_{\mathbf{n},c}\bar{\Phi}^{k}_c(\mathbf{n},t)
\Phi^{k}_c(\mathbf{n},t) +
\sum_{\mathbf{l},\alpha,c}\bar{\Phi}^{l}_c(\mathbf{n},\alpha,t)
\Phi^{l}_c(\mathbf{n},\alpha,t) 
\label{G:3}
\eeq
are invariant with respect to the gauge transformations (\ref{G:1} -\ref{G:2}).

The construction of a representation for the covariant derivative can 
be deduced from the continuum covariant derivative 
\beq
-i \pmb{\cal D}_{cc'} = - i\pmb{\nabla}\delta_{cc'} 
+  \mathbf{A}_d (\mathbf{x},t) \lambda^d_{cc'} 
\label{G:4}
\eeq
where $\lambda^d_{cc'}$ are the Gell-Mann matrices.  The projection of this 
operator on the wavelet basis gives
matrix elements of the form
\beq
-i \pmb{\cal D}^k_{\mathbf{m} \mathbf{n};cc'}
=-i \pmb{\nabla}^k_{\mathbf{m} \mathbf{n};cc'} + 
\sum_d\mathbf{A}^k_{\mathbf{m} \mathbf{n};d}(\mathbf{x},t)\lambda^d_{cc'}
\label{G:5}
\eeq

\beq
-i \pmb{\cal D}^{kl}_{\mathbf{m} \mathbf{n}\alpha ;cc'}
=-i \pmb{\nabla}^{kl}_{\mathbf{m} \mathbf{n}\alpha ;cc'}
+ \sum_d\mathbf{A}^{kl}_{\mathbf{m} \mathbf{n}\alpha ;d}
(t)\lambda^d_{cc'}
\label{G:6}
\eeq

\beq
-i\pmb{\cal D}^{lk}_{\mathbf{m}\alpha \mathbf{n};cc'}
=-i\pmb{\nabla}^{lk}_{\mathbf{m}\alpha \mathbf{n};cc'}
+\sum_d\mathbf{A}^{lk}_{\mathbf{m}\alpha \mathbf{n};d}
(t)\lambda^d_{cc'}
\label{G:7}
\eeq

\beq
-i\pmb{\cal D}^{jl}_{\mathbf{m}\alpha \mathbf{n}\beta;cc'}
=-i\pmb{\nabla}^{jl}_{\mathbf{m}\alpha \mathbf{n}\beta;cc'}
+\sum_d\mathbf{A}^{jl}_{\mathbf{m}\alpha \mathbf{n}\beta;d}
(t) \lambda^d_{cc'}
\label{G:8}
\eeq
where the multi-index quantities appearing in (\ref{G:5}-\ref{G:8})
are the matrix elements in the wavelet basis 
\beq
\pmb{\nabla}^k_{\mathbf{m} \mathbf{n};cc'} =
\int s_{\mathbf{m}}^k(\mathbf{x}) \pmb{\nabla} \delta_{cc'}
s_{\mathbf{n}}^k(\mathbf{x}) d\mathbf{x}, 
\label{G:9}
\eeq

\beq
\mathbf{A}^k_{\mathbf{m} \mathbf{n};d} =
\sum_l \int s_{\mathbf{m}}^k(\mathbf{x}) \mathbf{A}_d (\mathbf{x})  
s_{\mathbf{n}}^k(\mathbf{x}) d\mathbf{x}, 
\label{G:10}
\eeq

\beq
\pmb{\nabla}^{kl}_{\mathbf{m} \mathbf{n}\alpha ;cc'} =
\int s_{\mathbf{m}}^k(\mathbf{x}) \pmb{\nabla} \delta_{cc'}
w_{\mathbf{n}\alpha }^l(\mathbf{x}) d\mathbf{x}, 
\label{G:11}
\eeq

\beq
\mathbf{A}^{kl}_{\mathbf{m} \mathbf{n}\alpha;d} = 
\sum_l \int s_{\mathbf{m}}^k(\mathbf{x}) \mathbf{A}_d (\mathbf{x})  
w_{\mathbf{n}\alpha}^l(\mathbf{x}) d\mathbf{x}, 
\label{G:12}
\eeq

\beq
\pmb{\nabla}^{lk}_{\mathbf{m}\alpha \mathbf{n};cc'} =
\int w_{\mathbf{m}\alpha}^l(\mathbf{x}) \pmb{\nabla} \delta_{cc'}
s_{\mathbf{n}}^k(\mathbf{x}) d\mathbf{x}, 
\label{G:13}
\eeq

\beq
\mathbf{A}^{lk}_{\mathbf{m}\alpha \mathbf{n};d} =
\sum_l \int w_{\mathbf{m}\alpha}^l(\mathbf{x}) \mathbf{A}_d (\mathbf{x})
s_{\mathbf{n}}^k(\mathbf{x}) d\mathbf{x}, 
\label{G:14}
\eeq

\beq
\pmb{\nabla}^{jl}_{\mathbf{m}\alpha \mathbf{n}\beta;cc'} =
\int w_{\mathbf{m}\alpha}^j(\mathbf{x}) \pmb{\nabla} \delta_{cc'}
w_{\mathbf{n}\beta}^l(\mathbf{x}) d\mathbf{x}, 
\label{G:15}
\eeq

\beq
\mathbf{A}^{jl}_{\mathbf{m}\alpha \mathbf{n}\beta;d} =
\sum_l \int w_{\mathbf{m}\alpha}^j(\mathbf{x})\mathbf{A}_d (\mathbf{x})  
w_{\mathbf{n}\beta}^l(\mathbf{x}) d\mathbf{x}. 
\label{G:16}
\eeq

This derivative will transform covariantly: 
\beq
V^k_{cc'}(\mathbf{m},t)  \pmb{\cal D}^{k}_{\mathbf{m} \mathbf{n}\alpha ;cc'} =
\pmb{\cal D}^{k\prime }_{\mathbf{m} \mathbf{n} ;cc'} 
V^k_{cc'}(\mathbf{n}, t),
\label{G:17}
\eeq

\beq
V^k_{cc'}(\mathbf{m},t)  \pmb{\cal D}^{kl}_{\mathbf{m} \mathbf{n}\alpha ;cc'} =
\pmb{\cal D}^{kl\prime }_{\mathbf{m} \mathbf{n}\alpha ;cc'} 
V^l_{cc'}(\mathbf{n},\alpha, t),
\label{G:18}
\eeq

\beq
V^l_{cc'}(\mathbf{m},\alpha ,t)  \pmb{\cal D}^{lk}_{\mathbf{m}\alpha \mathbf{n};cc'} =
\pmb{\cal D}^{lk\prime }_{\mathbf{m} \alpha\mathbf{n} ;cc'} 
V^k_{cc'}(\mathbf{n},\alpha, t),
\label{G:19}
\eeq

\beq
V^j_{cc'}(\mathbf{n},\alpha,t)  \pmb{\cal D}^{jl}_{\mathbf{m}\alpha \mathbf{n}
\beta ;cc'} =
\pmb{\cal D}^{kl\prime }_{\mathbf{m}\alpha \mathbf{n}\beta ;cc'} 
V^l_{cc'}(\mathbf{n},\beta, t),
\label{G:20}
\eeq
provided the vector potential matrix transforms like 
\[
-i V^k_{cc'}(\mathbf{m}, t) \pmb{\nabla}^k_{\mathbf{m} \mathbf{n}} + 
\sum_{dc''}
V^k_{cc''}(\mathbf{m}, t) \mathbf{A}^k_{\mathbf{m} \mathbf{n};d}\lambda^d_{c''c} 
=
\]
\beq
-i \pmb{\nabla}^k_{\mathbf{m} \mathbf{n}}V^k_{cc'}(\mathbf{n}, t) + 
\sum_{dc''} \mathbf{A}^{\prime k}_{\mathbf{m} \mathbf{n};d}\lambda^d_{cc''}
V^k_{c''c}(\mathbf{n}, t)
\label{G:21}
\eeq
\[
-i V^k_{cc'}(\mathbf{m}, t) \pmb{\nabla}^{kl}_{\mathbf{m} \mathbf{n}\alpha} + 
\sum_{dc''}
V^k_{cc''}(\mathbf{m}, t) \mathbf{A}^{kl}_{\mathbf{m} \mathbf{n}\alpha;d}\lambda^d_{c''c} 
=
\]
\beq
-i \pmb{\nabla}^{kl}_{\mathbf{m} \mathbf{n}\alpha}V^l_{cc'}(\mathbf{n},\alpha,t) + 
\sum_{dc''}\mathbf{A}^{\prime kl}_{\mathbf{m} \mathbf{n};d}\lambda^d_{cc''}
V^l_{c''c}(\mathbf{n},\alpha t)
\label{G:22}
\eeq
\[
-i V^l_{cc'}(\mathbf{m},\alpha, t) \pmb{\nabla}^{lk}_{\mathbf{m}\alpha \mathbf{n}} + 
\sum_{dc''}
V^l_{cc''}(\mathbf{m},\alpha, t) \mathbf{A}^{lk}_{\mathbf{m} \alpha\mathbf{n} ;d}
\lambda^d_{c''c} 
=
\]
\beq
-i \pmb{\nabla}^{lk}_{\mathbf{m}\alpha \mathbf{n} } V^k_{cc'}(\mathbf{n}, t) + 
\sum_{dc''}\mathbf{A}^{\prime lk}_{\mathbf{m}\alpha \mathbf{n} ;d}\lambda^d_{cc''}
V^k_{c''c}(\mathbf{n}, t)
\label{G:23}
\eeq
\[
-i V^j_{cc'}(\mathbf{m},\alpha, t) \pmb{\nabla}^{jl}_{\mathbf{m}\alpha  
\mathbf{n}\beta} + 
\sum_{dc''}
V^j_{cc''}(\mathbf{m},\alpha, t) \mathbf{A}^{jl}_{\mathbf{m}\alpha 
\mathbf{n}\beta;d}
\lambda^d_{c''c} 
=
\]
\beq
-i \pmb{\nabla}^{jl}_{\mathbf{m} \alpha \mathbf{n}\beta}V^j_{cc'}(\mathbf{n}, 
\alpha,t) + 
\sum_{dc''}\mathbf{A}^{\prime jl}_{\mathbf{m} \alpha \mathbf{n}\beta;d}\lambda^d_{cc''}
V^l_{c''c}(\mathbf{n},\beta, t).
\label{G:24}
\eeq
With these transformation properties the quantities 
\beq
-i \sum_{\mathbf{m}\mathbf{n}} \bar{\Phi}^k_c(\mathbf{m}, t)
\pmb{\cal D}^k_{\mathbf{m} \mathbf{n};cc'}
\bar{\Phi}^k_c(\mathbf{n}, t),
\label{G:25}
\eeq

\beq
-i \sum_{\mathbf{m}\alpha \mathbf{n}} \bar{\Phi}^l_c(\mathbf{m},\alpha, t)
\pmb{\cal D}^{lk}_{\mathbf{m}\alpha \mathbf{n};cc'}
\bar{\Phi}^k_c(\mathbf{n}, t),
\label{G:26}
\eeq

\beq
-i \sum_{\mathbf{m}\mathbf{n}} \bar{\Phi}^k_c(\mathbf{m}, t)
\pmb{\cal D}^{kl}_{\mathbf{m} \mathbf{n}\alpha;cc'}
\bar{\Phi}^l_c(\mathbf{n},\alpha, t),
\label{G:27}
\eeq

\beq
-i \sum_{\mathbf{m}\mathbf{n}} \bar{\Phi}^{j}_c(\mathbf{m},\alpha, t)
\pmb{\cal D}^{jl}_{\mathbf{m}\alpha \mathbf{n}\beta;cc'}
\bar{\Phi}^l_c(\mathbf{n},\beta, t),
\label{G:28}
\eeq
are invariant.

Because these are invariant, matrix element by matrix element,
the invariance is  preserved by truncation.   A covariant field
strength tensor is obtained by taking the commutator of the 
covariant derivatives projected on different axes.   

Note that the generalization to 3+1 dimensions is a direct
extension of the three-dimensional results.  The interesting 
feature is that in the full theory the gauge invariance is implemented
by an infinite number of independent non-local gauge transformations
that act independently on each degree of freedom. 

\section{Summary and Conclusion}

In this paper we discussed some of the advantages of using the basis
generated by Daubechies scaling functions and wavelets to formulate
exact discretizations of local field theories.  We emphasized the
special properties of the basis and how these could be useful in
field-theoretic applications.  The most important properties are
that the basis functions have compact support, contain locally-finite
partitions of unity, and are related to fixed points of a renormalization
group equation.

The working assumption is that it is possible to make sense out of
quantum fields smeared with a class of test functions that are not
infinitely differentiable, but instead have a fractal character with a
limited amount of smoothness.  The justification for this is that the
integral of a product of Daubechies $K=3$ scaling functions over the
K\"all\'en-Lehmann representation of a mass $\mu$ two-point Wightman
function exists.  This justifies the use of wavelet smeared fields in
free-field theories and more generally in models with volume and
resolution cutoffs.

Given this assumption the wavelet basis leads to an exact
representation of the local field as an infinite linear combination of
smeared fields multiplied by compactly supported basis functions.  The
smeared field operators are operators rather than operator valued
distributions.  They generate a local algebra in the sense that in any
open set of spacetime there are operators associated with smearing
functions that have support entirely in that open set.

In the wavelet representation products of local field operators are
replaced by infinite sums of well-defined operators multiplied by
products of wavelet basis functions at different space-time points.  In
this representation singularities in the operator products at nearby
points are replaced by convergence questions.  For example, the local
composite fields in the Wilson-Zimmermann formulation of the operator
product expansion\cite{wilson} are recursively constructed by
identifying and ordering the most singular matrix elements as the
separation between points vanish.  In the wavelet representation these
matrix elements are represented by infinite sums of products
of basis functions at different points with well-defined expectation 
values of discrete field operators.  The identification of the singularity 
class of a given matrix elements depends on the asymptotic properties 
of the series as a function of the separation between points.

The discussions in this paper focused on fields smeared only over
wavelet basis functions in three dimensions at a fixed time.  This
representation is more convenient for dealing with the Hamiltonian
formulation of the dynamics as well as the treatment of the Poincar\'e
symmetry.  In this representation, if the basis is truncated to a
finite number of degrees of freedom (equivalent to a volume and
resolution cutoff), then the interactions in the Hamiltonain become
well-defined operators and one can in principle solve for the
truncated dynamics.  Because the basis functions are related to fixed
points of a renormalization group equation, Hamiltonians with fine
scale degrees of freedom are equal to identical Hamiltonians with
coarse scale degrees of freedom plus additional operators with
additional fine scale degrees of freedom.  Eliminating the fine scale
degrees of freedom, rescaling and adjusting the parameters of the
theory leads to a renormalization group transformation.  While the
calculations are not trivial, implementation of the decoupling using
similarity renormalization group method involves commutators of
discrete canonical fields and algebraically computable coefficients.
Methods for computing these coefficients were discussed in section five
and implemented in the appendix for the parts of the free Hamiltonain
that couple different scales.

Another important property of the wavelet basis is that the scaling 
functions on any fixed scale are up to an overall constant, a locally
finite,  compactly supported partition of unity.  When inserted in
the formal expressions for the Poincar\'e generators, expressed as
integrals of the energy momentum and angular momentum densities over 
a fixed time surface, the generator is decomposed into a sum of 
operators that act in different spatial volumes at a given time.
This allows one to exactly test the Poincar\'e commutation relations
in finite volumes.  These exact finite-volume components of the generators
can be expressed in terms of the discrete wavelet fields by replacing 
each field that appears in the generator by its expansion in wavelet 
smeared fields.  When these expansions are truncated, by eliminating
small-scale degrees of freedom, the commutation relations are violated 
because the products of the small-scale degrees of freedom couple to the 
large-scale degrees of freedom.  In the wavelet basis these correction
can be identified and their relative importance can be calculated. 
 
A final important property of the wavelet basis is the ability to 
reduce all of the quadratures that are needed in the theory to 
finite algebra.  This requires replacing conventional computational
techniques that depend on functions looking smooth on small scales by
new methods based on the renormalization group.

We also demonstrated that $SU(3)$ gauge invariance could be implemented
exactly in wavelet truncated theories.  While our discussion was
limited to some illustrative topics, there are a number of other
topics where the wavelet representation might have some
advantages. These include the wavelet representation of the operator
product expansion, wavelet representations of the Poincar\'e Lie algebra
in momentum space.

This work was supported in part by the U.S. Department of Energy, under
contract DE-FG02-86ER40286.  The authors would also like to 
thank Professor Robert Perry for useful discussions on the similarity 
renormalization group method and Professors Andreas Schaefer and 
Mikhail Altaisky 
for bringing our attention to some applications of wavelets in quantum 
field theory.

\section{Appendix - overlap integrals}

In this appendix we compute the overlap integrals
$D^k_{\mathbf{m}\mathbf{n}}$ that appear in the free-field Hamiltonian.  These
are the terms responsible for the coupling of the degrees of freedom
associated with different scales.

The general coefficients are products of the matrices below in
the $x$, $y$ and $z$ variables.  There are nine combinations.
\beq
D^k_{mn} = \int s^{k\prime}_m(x) s^{k\prime}_n(x) dx =
2^{2k} \int s' (x-m) s'(x-n) = 2^k D_{mn}   
\label{AP:1}
\eeq
\beq
D^{kl}_{mn} = \int s^{k\prime}_m(x) w^{l\prime}_n(x) dx =
2^{2(l+1)}\sum_{m'n'} H^{l+1-k}_{mm'}G_{nn'}D_{m'n'}
\label{AP:2}
\eeq
\beq
D^{jl}_{mn} = \int w^{j\prime}_m(x) w^{l\prime}_n(x) dx =
2^{2(l+1)}\sum_{m'n'} (GH^{l-j})_{mm'}G_{nn'} D_{m'n'} \qquad (l\geq j) .
\label{AP:3} 
\eeq 

The above expressions show that each of these integrals are linear combinations 
of the matrices   
\beq
D_{lm} = \int s_l'(x) s_m'(x) dx .
\label{AP:4}
\eeq
We can use translational invariance to write (\ref{AP:4}) equation as 
\beq
D_{lm} = D_{0,m-l} = \int s'(x) s_{m-l}'(x) dx .
\label{AP:5}
\eeq
Because of the support conditions on the scaling functions 
these vanish unless $\vert n-m\vert \leq 4$. 
It is difficult to get an inhomogeneous equation for $D_{lm}$
because the obvious choice, using the partition of unity (\ref{D:27}), 
satisfies
\beq
\sum_l l D_{0,m-l} = 0 
\label{AP:6}
\eeq
for any $m$.  This is because what remains is the integral of the derivative 
of a continuous function with compact support.  To get around 
this problem we use the 
partition of unity (\ref{A:15}) to get the relation
\beq
D_{0,m-l}= D_{lm} = \sum_n D_{nlm} = \sum_n \int s_n(x) s_l'(x) s_m'(x) dx .
\label{AP:7}
\eeq
This expresses the coefficients $D_{mn}$ as linear combinations of 
$D_{lmn}$.
Using the methods discussed in section five we find that 
coefficients $D_{nlm}= D_{0,l-n,m-n}$ satisfy the Homogeneous equations 
\beq
D_{0lm} = 4\sqrt{2} \sum_{n,k,j} H_{0n} H_{l,k+n}H_{m,j+n}D_{0kj}
\label{AP:8}
\eeq
and 
\beq
\sum_m D_{0lm} = 0 \qquad D_{0lm} - D_{0ml}=0 
\label{AP:9}
\eeq
and the inhomogeneous equation  
\beq
\sum_l l D_{nlm} = \int s_n (x)s'_m(x) dx = \Gamma_{nm}= \Gamma_{0,m-n} .
\label{AP:10}
\eeq
An independent subset of these equation can be solved for 
the non-zero  $D_{0lm}$s in terms of the $\Gamma_{0n}$.
The coefficients $\Gamma_{0n}$ satisfy the homogeneous equations 
\beq
\Gamma_{0l} = 2 \sum_{m,n } H_{0m}H_{l,n+m} \Gamma_{0n} 
\label{AP:11}
\eeq
and 
\beq
\sum_n n \Gamma_{0n} = 1 
\label{AP:12}
\eeq 
where we have used the partition of unity (\ref{D:27}).  These equations
can be solved for the non-zero $\Gamma_{0n}$ and the solutions can 
be used to calculate $D_{0lm}$ from which one can derive $D_{0m}$ 
using (\ref{AP:7}).  These can be used to calculate (\ref{AP:1}-\ref{AP:3})
and products of these quantities give the coefficients 
(\ref{C:35}-\ref{C:37}) for any combination of indicies.
The results of the calculation of the nine non-vanishing $D_{0m}$
are given in Table 2.

\begin{table} 
\centering
\begin{tabular}{|l|r|}
\hline
$D_{0-4}$ & $-5.357\times 10^{-3}$ \\ 
$D_{0-3}$ & $-1.143\times 10^{-1} $ \\
$D_{0-2}$& $ 8.762\times 10^{-1} $\\
$D_{0-1}$& $ -3.390\phantom{\times 10^{-11}}$\\
$D_{00}$& $ 5.268\phantom{\times 10^{-11}}$\\
$D_{01}$& $ -3.390\phantom{\times 10^{-11}}$\\
$D_{02}$& $ 8.762\times 10^{-1} $\\
$D_{03}$& $ -1.143\times 10^{-1} $\\
$D_{04}$& $ -5.357\times 10^{-3}$ \\
\hline
\end{tabular}
\caption{}
\end{table}
 
The results of the calculation of the non zero $\Gamma_{0n}$
and  $D_{0mn}$  
are given in tables 3 and 4.

\begin{table}
\centering
\begin{tabular}{|l|r|}
\hline
$\Gamma_{0-4}$& $-3.424658\times 10^{-4} $ \\
$\Gamma_{0-3}$& $-1.461187\times 10^{-2} $ \\
$\Gamma_{0-2}$& $ 1.452055\times 10^{-1}  $\\
$\Gamma_{0-1}$& $-7.452055\times 10^{-1}  $\\
$\Gamma_{00}$& $ -5.116622\times 10^{-16}  $\\
$\Gamma_{01}$& $  7.452055\times 10^{-1}  $\\
$\Gamma_{02}$& $ -1.452055\times 10^{-1} $\\
$\Gamma_{03}$& $  1.461187\times 10^{-2}  $\\
$\Gamma_{04}$& $  3.424658\times 10^{-4}  $\\
\hline
\end{tabular}
\caption{}
\end{table}

\begin{table}
\centering
\begin{tabular}{|l|r|l|r|l|r|}
\hline
$D_{0-4-4 }$ & $ 4.056756\times 10^{-5} $ &$D_{0-1-2}$& $-6.544856\times 10^{-1\
} $  &$D_{012}$& $ 1.758631\times 10^{-1} $  \\
$D_{0-4-3}$  & $ 1.620980\times 10^{-4} $ &$D_{0-1-1}$& $ 2.323493\phantom{\times 10^{-111}} $  &$D_{013}\
$& $ 1.299066\times 10^{-2} $  \\
$D_{0-4-2}$  & $-6.227505\times 10^{-4} $ &$D_{0-1 0}$& $-2.071142\phantom{\times 10^{-111}} $  &$D_{014}\
$& $-8.819594\times 10^{-5} $  \\
$D_{0-4-1}$  & $ 9.394026\times 10^{-4} $ &$D_{0-1 1}$& $ 3.284401\times 10^{-1\
} $  &$D_{02-2}$& $ 1.773552\times 10^{-3} $ \\
$D_{0-4 0}$  & $-5.193176\times 10^{-4} $ &$D_{0-1 2}$& $ 3.304668\times 10^{-2\
} $  &$D_{02-1}$& $ 3.304668\times 10^{-2} $ \\
$D_{0-3 -4}$ & $ 1.620980\times 10^{-4} $ &$D_{0-1 3}$& $-3.892382\times 10^{-4\
} $  &$D_{020}$& $-8.023148\times 10^{-2} $  \\
$D_{0-3 -3}$ & $ 1.782152\times 10^{-2} $ &$D_{00-4}$&  $-5.193176\times 10^{-4\
} $  &$D_{021}$&  $ 1.758631\times 10^{-1} $ \\
$D_{0-3 -2}$ & $-4.290543\times 10^{-2} $ &$D_{00-3 }$& $-8.962795\times 10^{-3\
} $  &$D_{022 }$& $-1.066658\times 10^{-1} $ \\
$D_{0-3 -1}$ & $ 4.009720\times 10^{-2} $ &$D_{00-2}$&  $ 5.753018\times 10^{-1\
} $  &$D_{023}$&  $-2.400100\times 10^{-2} $ \\
$D_{0-3 0}$  & $-8.962795\times 10^{-3} $ &$D_{00-1}$&  $-2.071142\phantom{\times 10^{-111}} $  &$D_{024}\
$&  $ 2.149391\times 10^{-4} $ \\
$D_{0-3 1}$  & $-6.212589\times 10^{-3} $ &$D_{000}$&   $ 2.364229\phantom{\times 10^{-111}} $  &$D_{03-1\
}$& $-3.892382\times 10^{-4} $ \\
$D_{0-2 -4}$ & $-6.227505\times 10^{-4} $ &$D_{001}$&   $-7.734980\times 10^{-1\
} $  &$D_{030}$&  $-5.168265\times 10^{-3} $ \\
$D_{0-2 -3}$ & $-4.290543\times 10^{-2} $ &$D_{002}$&   $-8.023148\times 10^{-2\
} $  &$D_{031}$&  $ 1.299066\times 10^{-2} $ \\
$D_{0-2 -2}$ & $ 2.549910\times 10^{-1} $ &$D_{003}$&   $-5.168265\times 10^{-3\
} $  &$D_{032}$&  $-2.400100\times 10^{-2} $ \\
$D_{0-2 -1}$ & $-6.544856\times 10^{-1} $ &$D_{004}$&   $-9.550006\times 10^{-6\
} $  &$D_{033}$&  $ 1.703740\times 10^{-2} $ \\
$D_{0-2 0}$  & $ 5.753018\times 10^{-1} $ &$D_{01-3}$&  $-6.212589\times 10^{-3\
} $  &$D_{034}$&  $-4.695561\times 10^{-4} $ \\
$D_{0-2 1}$ & $-1.340525\times 10^{-1} $ &$D_{01-2 }$& $-1.340525\times 10^{-1}\
 $  &$D_{040 }$& $-9.550006\times 10^{-6} $ \\
$D_{0-2 2}$  & $ 1.773552\times 10^{-3} $ &$D_{01-1}$ & $ 3.284401\times 10^{-1\
} $  &$D_{041}$ & $-8.819594\times 10^{-5} $ \\
$D_{0-1-4 }$ & $ 9.394026\times 10^{-4} $ &$D_{010}$&   $-7.734980\times 10^{-1\
} $  &$D_{042}$&  $ 2.149391\times 10^{-4} $ \\
$D_{0-1-3}$  & $ 4.009720\times 10^{-2} $ &$D_{011}$&   $ 3.965574\times 10^{-1\
} $  &$D_{043}$&  $-4.695561\times 10^{-4} $ \\
$$  & $ $ &$ $ &   $ $  &$D_{044}$&  $ 3.523629\times 10^{-4} $ \\
\hline
\end{tabular}
\caption{}
\end{table}


\begin{thebibliography}{}

\bibitem{daubechies:1988} I. Daubechies, 
Comm. Pure Appl. Math. {\bf 41},909(1988).

\bibitem{daubechies}I. Daubechies, {\it Ten Lectures on Wavelets},
CBMS-NSF Regional Conference Series in Applied Mathematics,
1992.  

\bibitem{kaiser} G. Kaiser, {\it A Friendly Guide to Wavelets}, Birkhauser 1994.

\bibitem{resnikoff}Howard L. Resnikoff and Raymond O. Wells, Jr..
{\it Wavelet Analysis}, Springer, 1998.

\bibitem{jorgensen} O. Bratteli and P. Jorgensen, {Wavelets through A
Looking Glass - The World of the Spectrum}, 
Birkh\"auser, 2002.

\bibitem{wavelets}B. M. Kessler, G. L. Payne, W. N. Polyzou,
Wavelet Notes, arXiv:nucl-th/0305025v2, 2003. 

\bibitem{best:1994} Christoph Best, Andreas Schaefer,
''Variational description of statistical field theories using Daubechies' 
wavelets'', arXiv: hep-lat/9402012, 1994.

\bibitem{best:2000} Christoph Best, 
Nucl. Phys. Proc. Suppl. {\bf 83},848(2000).

\bibitem{Ismail1:2003} Ahmed E. Ismail, Gregory C. Rutledge, 
and George Stephanopoulos,  
J. Chem. Phys. {\bf 118},4414(2003).

\bibitem{Ismail2:2003} Ahmed E. Ismail, Gregory C. Rutledge, 
and George Stephanopoulos,  
J. Chem. Phys. {\bf 118},4424(2003).

\bibitem{federbush:1995} P. Federbush,
Prog. Theor. Phys. {\bf 94},1135(1995).

\bibitem{Battle:1999} Guy Battle, {\it Wavelets and Renormalization}, 
Series in Approximations and Decompositions, Volume 10, 
World Scientific, 1999.

 \bibitem{altaisky:2007} Mikhail V. Altaisky,
SIGMA {\bf 3},105(2007).

\bibitem{albeverio:2009} S. Albeverio, Mikhail V. Altaisky,
``A remark on gauge invariance in wavelet-based quantum field theory''
arXiv:0901.2806v2,2009.

\bibitem{altaisky:2010} Mikhail V. Altaisky,
Phys. Rev. D {\bf 81},125003(2010).

\bibitem{perry}
E. Anderson, S.K. Bogner, R.J. Furnstahl, E.D. Jurgenson, R.J. Perry, A. Schwenk,
Block Diagonalization using SRG Flow Equations,
arXiv:0801.1098,2008.

\bibitem{okobu}
Okubo S., Prog. Theor. Phys. {\bf 12},603(1954).

\bibitem{glockle} 
Gl\"ockle W., Mueller L., Phys. Rev. C{\bf 23},1183(1981).

\bibitem{wilson} Kenneth G. Wilson and Wolfhart Zimmermann,
Comm. Math. Phys. {\bf 24},87(1972). 



\end{thebibliography}
\end{document}